\documentclass[12pt]{article}

\usepackage{amsmath}
\usepackage{graphicx,epsf}
\usepackage{natbib}
\usepackage[font=scriptsize,labelfont=bf]{caption}
\usepackage{graphicx,multicol}
\usepackage{color}
\usepackage{booktabs}  
\usepackage{float}
\usepackage[section]{placeins}
\usepackage[caption = false]{subfig}
\usepackage{multirow}
\usepackage[toc,page]{appendix}
\usepackage[ruled]{algorithm2e}
\usepackage{pdflscape}

\RequirePackage[OT1]{fontenc}
\RequirePackage{amsthm,amsmath,amssymb}
\RequirePackage{natbib}

\definecolor{micolo}{rgb} {0.0, 0.26, 0.15}

\DeclareMathOperator*{\argmax}{argmax}
\usepackage{natbib}

\newtheorem{remark}{Remark}

\numberwithin{equation}{section}
\theoremstyle{plain}

\newcommand*{\bigCI}{
  \mathrel{\text{
    {\rotatebox[origin=c]{90}{\resizebox{2.25ex}{1.65ex}{$\vDash$}}}
  }}
}

\newcommand\vari {\mbox{\sc var}}

\newcommand\cova {\mbox{\sc cov}}
\newcommand\corr {\mbox{\sc corr}}
\newcommand\supp {\mbox{\sc supp}}

\begin{document}

\def\spacingset#1{\renewcommand{\baselinestretch}%
{#1}\small\normalsize} \spacingset{1}


\author{
  \small Ginette LAFIT, Francisco J. NOGALES, Marcelo RUIZ and Ruben H. ZAMAR 
	}

\title{\bf A Stepwise Approach for High-Dimensional Gaussian Graphical Models}
\date{}

  \maketitle
	
	\makeatletter{\renewcommand*{\@makefnmark}{}
{\footnotetext{ \scriptsize Ginette Lafit, Postdoctoral research fellow, Research Group of Quantitative Psychology and Individual Differences, KU Leuven–University of Leuven, Leuven, Belgium (E-mail:  ginette.lafit@kuleuven.be), Francisco J. Nogales is  Professor, Department of Statistics and UC3M-BS Institute of Financial Big Data, Universidad Carlos III de Madrid, Espa\~na (E-mail: fcojavier.nogales@uc3m.es),  Ruben H. Zamar is Professor, Department of Statistics, University of British
Columbia, 3182 Earth Sciences Building, 2207 Main Mall, Vancouver, BC  V6T 1Z4, Canada (Email:
ruben@stat.ubc.ca) and Marcelo Ruiz is Professor, Departamento de Matem\'atica, FCEFQyNat, Universidad Nacional de R\'io Cuarto, C\'ordoba, Argentina (E-mail: mruiz@exa.unrc.edu.ar). }}\makeatother}

\begin{abstract}

We present a stepwise approach to estimate  high dimensional Gaussian graphical models . We exploit the relation between the partial correlation coefficients and the distribution of the prediction errors, and parametrize  the model in terms of the Pearson correlation coefficients between the prediction errors of the nodes' best linear predictors.  We propose a novel stepwise algorithm for detecting pairs of conditionally dependent variables. We show that the  proposed algorithm outperforms existing methods such as the graphical lasso and CLIME in simulation studies and real life applications. In our comparison we report different performance measures that look at different desirable features of  the recovered  graph and consider  several  model settings.

\end{abstract}


\vfill

\newpage
\spacingset{1.45}  
\section{Introduction}\label{sec_1}

High-dimensional {\it Gaussian graphical models} (GGM) are widely used in practice to represent the linear dependency between variables. The  underlying idea in GGM is to measure linear dependencies by estimating partial correlations to infer whether there is an association between a given pair of variables, conditionally on the remaining ones. Moreover, there is a close relation between the nonzero partial correlation coefficients and the nonzero entries in the inverse of the covariance matrix.
 Covariance selection procedures  take advantage of this fact  to estimate   the GGM conditional dependence structure given a sample \citep{dempster1972covariance, lauritzen1996graphical,edwards2000introduction}.

When the dimension $p$ is larger than the number $n$ of observations,  the sample covariance matrix $S$ is not  invertible and the maximum likehood estimate (MLE)  of $\boldsymbol{\Sigma}$ does not exist.  When    $p/n \leq 1$,  but close to $1$, $S$ is invertible but  ill-conditioned, increasing the estimation error \citep{ledoit2004well}.   To deal with this problem, several covariance selection procedures have been proposed based on the assumption that the inverse of the covariance matrix, $\Omega$, called  {\it precision matrix}, is sparse.   

We present an approach to perform covariance selection in a high dimensional  GGM based on a forward-backward algorithm called \textit{graphical stepwise} (GS).  Our procedure takes advantage of the  relation  between the partial correlation  and the Pearson correlation coefficient of the residuals. 

Existing methods to estimate the GGM can be classified in three classes:  nodewise regression methods, maximum likelihood methods and limited order partial correlations methods. The nodewise regression method was proposed by \cite{meinshausen2006high}. This method estimates a lasso regression for each node in the graph. See for example  \cite{peng2009partial}, \cite{yuan2010high}, \cite{liu2012tiger}, \cite{zhou2011high}  and \cite{ren2015asymptotic}.
Penalized likelihood methods include \cite{yuan2007model}, \cite{banerjee2008model},    \cite{friedman2008sparse}, \cite{johnson2011high} and    \cite{ravikumar2011high} among others.   \cite{cai2011constrained} propose an estimator called CLIME that estimates precision matrices by solving the dual of an $\ell_1$ penalized maximum likelihood problem.  
Limited order partial correlation procedures use lower order partial correlations to test for conditional independence relations. See \cite{spirtes2000causation}, \cite{kalisch2007estimating}, \cite{rutimann2009high},  \cite{liang2015equivalent} and \cite{huang2016}.

 The rest of the article is organized as follows. Section \ref{section_2} introduces the stepwise approach along with some notation.  Section \ref{section_3}  gives   simulations results and a real data example.  Section \ref{section_4} presents some concluding remarks. The Appendix shows a detailed description of the crossvalidation procedure used to determine  the required parameters in our stepwise algorithm and gives   some additional  results from our simulation study.

\section{Stepwise  Approach to Covariance Selection}
\label{section_2}
\subsection{Definitions and Notation}\label{sec_1a}

In this section we review some definitions and technical concepts  needed later on.  Let $\mathcal{G}=(V,E)$  be a graph where $V\neq \emptyset$ is the set of nodes or vertices  and $E\subseteq V\times V= V^2$ is the set of edges. For simplicity we assume  that $V=\{1,\ldots,p \}$.  We assume that the graph $\mathcal{G}$ is  undirected, that is,   $ (i,j)\in E$ if and only if $(j,i)\in E$.  Two nodes $i$ and $j$ are called connected, adjacent or neighbors if $(i,j) \in E$.

A  {\it graphical model} (GM) is a graph   such that $V$ indexes a set of variables $\{ X_1,\ldots,X_p\}$ and $E$ is defined by:
\begin{equation} \label{pmp}
 (i,j) \notin E \text{ if and only if } X_i \bigCI X_j \mid X_{V \setminus \{i,j\}.
}\end{equation}
Here $ \bigCI$ denotes  {\it conditional independence}.

Given a node $i\in V$,   its neighborhood $\mathcal{A}_i$ is defined as
\begin{equation} \label{eq2-1}
\mathcal{A}_i = \{l \in V \setminus \{i\}: (i,l) \in E 	\}.
\end{equation}
Notice that $\mathcal{A}_i$ gives the nodes directly connected with  $i$ and therefore a GM can be effectively described by giving the system of neighborhoods $\displaystyle \left\{\mathcal{A}_i\right\}_{i = 1}^{p}$.

 We further assume that  $\displaystyle  (X_1,\ldots,X_p)^{\top} \sim \text{N}(\boldsymbol{0}, \boldsymbol{\Sigma})$, where $\boldsymbol{\Sigma}=(\sigma_{ij})_{i,j=1 \ldots,p}$ is a  positive-definite covariance matrix. 
In  this case the graph is called a {\it Gaussian graphical model} (GGM). 
The matrix  $\boldsymbol{\Omega}=(\omega_{ij})_{i,j=1 \ldots,p}=\boldsymbol{\Sigma}^{-1}$ is called {\it precision matrix}.

 There exists an extensive  literature on GM and GGM.  For a detailed treatment of the theory see for instance   \cite{lauritzen1996graphical}, \cite{edwards2000introduction}, and \cite{buhlmann2011statistics}.

\subsection{Conditional dependence in a GGM} \label{cdep}

In a GGM  the set of edges $E$ represents the conditional dependence structure  of the vector $(X_1,\ldots, X_p)$.  To represent this dependence structure as  a statistical model it is convenient to find a parametrization for $E$.

In this subsection we introduce a convenient parametrization of $E$  using well known  results from classical multivariate  analysis.  For an  exhaustive treatment of these results see, for instance,  \citet{andersonmult}, \citet{cramer}, \citet{lauritzen1996graphical} and \citet{eaton2007}.

Given a subset  $\mathcal{A}$ of $V$,  $\mathbf{X}_{\mathcal{A}}$ denotes the vector of variables with subscripts in $\mathcal{A}$  in increasing order.  For a given pair of nodes $(i,l)$,   set   $\mathbf{X}_{1}^{\top}=\left(  X_{i},X_{l}\right)  $,  $\mathbf{X}_{2}^{\top}=\mathbf{X}_{V\backslash\left\{  i,l\right\}  } $  and  $ \mathbf{X}=\left(\mathbf{X}_{1}^{\top},\mathbf{X}_{2}^{\top} \right)^{\top}$.
Note that $ \mathbf{X}$ has multivariate normal distribution with mean $ \mathbf{0}$ and covariance matrix
\begin{equation} \label{matpart}
\begin{pmatrix}
\Sigma_{11} & \Sigma_{12}\\
\Sigma_{21} & \Sigma_{22}
\end{pmatrix}
\end{equation}
such that $\Sigma_{11}$   has dimension $2\times 2$,   $\Sigma_{12}$ has dimension $2\times (p-2)$ and so on. The matrix in \eqref{matpart} is a partition of a permutation of the original covariance matrix $\Sigma$, and will be also denoted by  $\Sigma$, after a small abuse of notation.

Moreover, we set
\begin{align*}
\Omega =
\begin{pmatrix}
\Sigma_{11} & \Sigma_{12}\\
\Sigma_{21} & \Sigma_{22}
\end{pmatrix}^{-1}   =
\begin{pmatrix}
\Omega_{11} & \Omega_{12}\\
\Omega_{21} & \Omega_{22}
\end{pmatrix}.
\end{align*}
Then,  by (B.2) of \citet{lauritzen1996graphical}, the blocks $\Omega_{i,j}$ can be written explicitly in terms of  $\Sigma_{i,j}$  and  $\Sigma_{i,j}^{-1}$. In particular

$\Omega_{11}    =\left(  \Sigma_{11}-\Sigma_{12}\Sigma_{22}^{-1}\Sigma_{21}\right)  ^{-1}$ 
where 
\[
\Omega_{11}=
\begin{pmatrix}
\omega_{ii} & \omega_{il}\\
\omega_{li} & \omega_{ll}
\end{pmatrix}
\]
is the submatrix of $\Omega$ (with rows $i$ and $l$ and columns $i$ and $l$).  Hence,
\begin{eqnarray} \label{alfin}
\cova \left(  \mathbf{X}_{1}|\mathbf{X}_{2}\right)  &=& \Sigma_{11}-\Sigma
_{12}\Sigma_{22}^{-1}\Sigma_{21}\\ \nonumber
  &=& \Omega_{11}^{-1} \\  \nonumber
 &=&  \frac{1}{  \omega_{ii}  \omega_{ll}- \omega_{il} \omega_{li} }
 \begin{pmatrix}
\omega_{ll} & -\omega_{il}\\
- \omega_{li} & \omega_{ii}
\end{pmatrix}
\end{eqnarray}
and, in consequence,  the partial correlation between $X_i$ and $X_l$ can be expressed as
\begin{equation} \label{eq2-3}
 \corr  \left( X_i,X_l |\mathbf{X}_{V\backslash\left\{  i,l\right\}  }  \right) = - \frac{\omega_{il}}{ \sqrt{\omega_{ii}\omega_{ll} }}.
\end{equation}
This gives the standard parametrization of $E$ in terms of  the support of the precision matrix
\begin{eqnarray}\label{suppomega}
 \supp\left( \Omega \right)= \{(i,l) \in  V^2:\, i\neq l, \omega_{i,l} \neq 0 \}.
\end{eqnarray}

We now  introduce another parametrization of $E$, which we need to define and implement our proposed method. We consider  the regression error for the regression of  $\mathbf{X}_{1}$ on $\mathbf{X}_{2}$,
   $$\displaystyle  \boldsymbol{\varepsilon} =  \mathbf{X}_{1} - \widehat{\mathbf{X}}_{1}=\mathbf{X}_{1} - \boldsymbol{\beta}^{\top}    \mathbf{X}_{2} $$  and let $\varepsilon_i$ and $\varepsilon_l$  denote the entries of $\boldsymbol{\varepsilon}$ (i.e. $\boldsymbol{\varepsilon}^{\top}=(\varepsilon_i, \varepsilon_l)$). 
The regression error $\boldsymbol{\varepsilon}$ is independent of $ \widehat{\mathbf{X}}_{1}$ and has normal distribution with mean $\boldsymbol{0}$ and covariance matrix  $\Psi_{11} $ with elements denoted by
\begin{align}
\Psi_{11}  &  =
\begin{pmatrix}
\psi_{ii} & \psi_{il}\\
\psi_{li} & \psi_{ll}
\end{pmatrix}.
\end{align}
A straightforward calculation shows that

\begin{align*}
\Psi_{11}  &  =   \cova \left(  \mathbf{X}_{1}\right)  +\cova\left(  \widehat{\mathbf{X}}
_{1}\right)  -2  \cova \left(  \mathbf{X}_{1},\widehat{\mathbf{X}}_{1}\right)
\nonumber\\
& \nonumber\\
&  =\Sigma_{11}+\Sigma_{12}\Sigma_{22}^{-1}\Sigma_{22}\Sigma_{22}^{-1}
\Sigma_{21}-2\Sigma_{12}\Sigma_{22}^{-1}\Sigma_{21}\nonumber\\
& \nonumber\\
&  =\Sigma_{11}-\Sigma_{12}\Sigma_{22}^{-1}\Sigma_{21}=\Omega_{11}^{-1}.
\label{eq4}
\end{align*}
See \citet[ Section 23.4]{cramer}.

Therefore, by this equality,  \eqref{alfin}  and \eqref{eq2-3}, the partial correlation coefficient and the conditional correlation are equal
\begin{eqnarray}
\rho_{il\cdot V\backslash\left\{  i,l\right\}  } =
 \corr  \left( X_i,X_l | \mathbf{X}_{V\backslash\left\{  i,l\right\}  } \right) \label{eq2-3bis}
= \frac{\psi_{il}}{\sqrt{\psi_{ii}\psi_{ll}}}. \nonumber
\end{eqnarray}
Summarizing,   the problem of determining the conditional dependence structure in a GGM (represented by $E$)  is equivalent to finding the pairs of nodes of $V$ that belong to the set
\begin{eqnarray}\label{supppartial}
 \{(i,l) \in  V^2:\, i\neq l, \psi_{i,l} \neq 0 \}
\end{eqnarray}
which is equal to  the support of the precision matrix, $\supp\left( \Omega \right)$, defined by  \eqref{suppomega}.

\begin{remark}

 As noticed above, under normality, partial and conditional correlation are the same. However,  in general  they are different concepts  \citep{lawrance}.

\end{remark}

\begin{remark} \label{Rem2}
 Let $ \beta_{i,l}$ be the regression coefficient of  $X_l$ in the regression of  $X_i$ versus $\mathbf{X}_{V\backslash\left\{  i\right\} }$ 
 and, similarly  let $\beta_{l,i}$ be  the regression coefficient of   $X_i$ in the regression of  $X_l$ versus $\mathbf{X}_{V\backslash\left\{  i\right\} }$. Then it follows that $\displaystyle  \rho_{il\cdot V\backslash\left\{  i,l\right\}  }=  \text{sign}  \left( \beta_{l,i}  \right) \sqrt{  \beta_{l,i}\beta_{i,l}    }$. This allows for another popular parametrization for $E$. Moreover, let $\epsilon_i$ be the error term  in the regression of the $i^{\text{th}}$ variable on the remaining ones. Then by Lemma 1 in \cite{peng2009partial} we have that $\cova( \epsilon_i,\epsilon_l)=\omega_{il}/\omega_{ii}\omega_{ll}$ and
$\vari( \epsilon_i)=1/\omega_{ii}$.
   \end{remark}

\subsection{The Stepwise Algorithm \bigskip}

Conditionally on its neighbors,  $X_i$ is independent of all the other variables. Formally, for all $i$, 
\begin{equation} \label{pairneigh}
\text{if  } l \notin {\cal A}_i\ \text{and }  l\neq i  \text{ then }  X_{i} \bigCI  X_{l} | \mathbf {X}_{{ \mathcal A_i}}.
\end{equation}
Therefore, given a system of neighborhoods $\left\{
\mathcal{A}_{i}\right\}  _{i=1}^{p}$ and $l\notin\mathcal{A}_{i}$ (and
so $i\notin\mathcal{A}_{l}$), the partial correlation between
$X_{i}$ and $X_{l}$ can be obtained by the following procedure: (i)  regress $X_{i}$ on $\mathbf{X}_{\mathcal{A}_{i}}$ and compute the regression
residual $\varepsilon_{i}$;  regress $X_{l}$ on $\mathbf{X}_{\mathcal{A}_{l}}$ and compute the regression
residual $\varepsilon_{l}$; (ii) calculate the Pearson correlation between $\varepsilon_{i}$ and $\varepsilon
_{l}.$ 

This reasoning motivates the  graphical stepwise algorithm (GSA). It begins with the family of empty neighborhoods, $\hat{\mathcal{A}}_j^{(0)} = \emptyset$ for each $j \in V$.  There are two basic steps, the forward and the backward steps.  In the forward step, the algorithm adds a new edge $(j_{0},l_{0})$ if the largest absolute empirical partial correlation between  the variables $X_{j_{0}}, X_{l_{0}}$ is above the given   threshold $\alpha_{f}$. In the backward step  the algorithm deletes an edge $(j_{0},l_{0})$ if the smallestt absolute empirical partial correlation between  the variables $X_{j_{0}}, X_{l_{0}}$ is below the given   threshold $\alpha_{b}$.
A step by step description of GSA is as follows: 
\medskip

{\small
\noindent \textbf{Graphical Stepwise Algorithm}

\begin{itemize}

\item[]  \textbf{Input:} the (centered) data $\left\{  \mathbf{x}_{1},...,\mathbf{x}_{n}\right\}
, $ and the forward and backward thresholds $\alpha_{f}$ \ and $\alpha_{b}.$

\item [] \textbf{Initialization}.  $k=0$: set $\widehat{\mathcal{A}}_{1}^{0}=\widehat{\mathcal{A}}_{2}
^{0}=\cdots=\widehat{\mathcal{A}}_{p}^{0}=\phi$. 

\item [] \textbf{Iteration  Step. } Given $\widehat{\mathcal{A}}_{1}^{k},\widehat{\mathcal{A}}_{2}^{k},...,\widehat{\mathcal{A}}_{p}^{k}$ \ we
compute $\widehat{\mathcal{A}}_{1}^{k+1},\mathcal{A}_{2}^{k+1},...,\widehat{\mathcal{A}}_{p}
^{k+1}$ \ as follows.

\begin{enumerate}
\item [] \textbf{Forward. }  For each $j=1,...,p$ \ do the following.

For each $l\notin \widehat{\mathcal{A}}_{j}^{k}$ \ calculate the partial correlations $f_{jl}^{k}$ as follows.

\begin{enumerate}
\item Regress the $j^{th}$ variable on the variables with subscript in the
set $\widehat{\mathcal{A}}_{j}^{k}$ \ and compute the regression residuals
$\mathbf{e}_{j}^{k}=\left(  e_{1j}^{k},e_{2j}^{k},...,e_{nj}^{k}\right).$

\item Regress the $l^{th}$ variables on the variables with subscript in the
set $\widehat{\mathcal{A}}_{l}^{k}$ \ and compute the regression residuals
$\mathbf{e}_{l}^{k}=\left(  e_{1l}^{k},e_{2l}^{k},...,e_{nl}^{k}\right).$

\item Obtain the partial correlation $f_{jl}^{k}$ by calculating the Pearson correlation  between $\mathbf{e}_{j}
^{k}$ and $\mathbf{e}_{l}^{k}.$ 
\end{enumerate}

If
\[
\max_{l\notin\widehat{\mathcal{A}}_{j}^{k},j\in V}\left\vert f_{jl}^{k}\right\vert
=\left\vert f_{j_{0}l_{0}}^{k}\right\vert \geq\alpha_{f}
\]
set $\widehat{\mathcal{A}}_{j_{0}}^{k+1}=\widehat{\mathcal{A}}_{j_{0}}^{k}\cup\left\{
l_{0}\right\}  ,$ $\widehat{\mathcal{A}}_{l_{0}}^{k+1}=\widehat{\mathcal{A}}_{l_{0}}^{k}
\cup\left\{  j_{0}\right\}  ,$ $\widehat{\mathcal{A}}_{l}^{k+1}=\widehat{\mathcal{A}}_{l}^{k}$
\ for $l\neq j_{0},l_{0}$ \

If
\[
\max\left\vert f_{jl}^{k}\right\vert =\left\vert f_{j_{0}l_{0}}^{k}\right\vert
<\alpha_{f}, \text{ stop}.
\]

\item []\textbf{Backward. }For each $j=1,...,p$ \ do the following.

For each $l\in\widehat{\mathcal{A}}_{j}^{k+1}$ \ calculate the partial correlation $b_{jl}^{k}$ as follows.

\begin{enumerate}
\item Regress the $j^{th}$ variables on the variables with subscript in the
set $\widehat{\mathcal{A}}_{j}^{k+1}\backslash\left\{  l\right\}  $ \ and compute the
regression residuals $\mathbf{r}_{j}^{k}=\left(  r_{1j}^{k},r_{2j}
^{k},...,r_{nj}^{k}\right)  .$

\item Regress the $l^{th}$ variable on the variables with subscript in the
set $\widehat{\mathcal{A}}_{l}^{k+1}\backslash\left\{  j\right\}  $ \ and compute the
regression residuals $\mathbf{r}_{l}^{k}=\left(  r_{1l}^{k},r_{2l}
^{k},...,r_{nl}^{k}\right)  .$

\item Compute the partial correlation $b_{jl}^{k}$ by calculating the Pearson correlation between $\mathbf{r}_{j}
^{k}$ and $\mathbf{r}_{l}^{k}.$ 
\end{enumerate}
If
\[
\min_{l\in\widehat{\mathcal{A}}_{j}^{k},j\in V}\left\vert b_{jl}^{k}\right\vert
=\left\vert b_{j_{0}l_{0}}^{k}\right\vert \leq\alpha_{b}
\]
set $\widehat{\mathcal{A}}_{j_{0}}^{k+1}\rightarrow\widehat{\mathcal{A}}_{j_{0}}^{k+1}
\backslash\left\{  l_{0}\right\}  ,$ $\widehat{\mathcal{A}}_{l_{0}}^{k+1}\rightarrow
\widehat{\mathcal{A}}_{l_{0}}^{k+1}\backslash\left\{  j_{0}\right\}$. 
\end{enumerate}

\bigskip

\item[] {\bf Output}
\begin{enumerate}  
\item A collection of estimated neighborhoods $\widehat{\mathcal{A}}_{j}$, $j=1,\ldots,p$.
\item The set of estimated edges $\widehat{E}=\left\{(i,l)\in V^2: i\in \widehat{\mathcal{A}}_{l} \right\}$.
\item An estimate of $\boldsymbol{\Omega }$, $\widehat{\boldsymbol{\Omega }}=\left( \widehat{\omega }
_{il}\right) _{i,l=1}^{p}$ with $\widehat{\omega }_{il}$ defined as follow:
in the case $i=l,$  $\widehat{\omega }_{ii}=n/(\boldsymbol{e}_{i}^{T}%
\boldsymbol{e}_{i})$ for $i=1,...,p,$ where $\boldsymbol{e}_{i}$ is the
vector of the prediction errors in the regression of the $i^{\text{th}}$
variable on $\boldsymbol{X}_{\widehat{\mathcal{A}}_{i}}.$ In the case $i\neq
l$ \ we must distinguish two cases, if $l\notin \widehat{\mathcal{A}}_{i}$
then $\widehat{\omega }_{il}=0,$ otherwise $\widehat{\omega }_{il}=n\left( 
\boldsymbol{e}_{i}^{T}\boldsymbol{e}_{l}\right) /\left[ \left( \boldsymbol{e}%
_{i}^{T}\boldsymbol{e}_{i}\right) \left( \boldsymbol{e}_{l}^{T}\boldsymbol{e}%
_{l}\right) \right]$ (see Remark \ref{Rem2}). 
\end{enumerate}
\end{itemize}
}

\subsection{ Thresholds selection     by cross-validation} \label{cross}

Let   $\boldsymbol{X}$ be the $n\times p$ matrix with  rows $\mathbf{x}_{i}=\left(x_{i1},\ldots,x_{ip} \right)$, $i=1,\ldots,n$,  corresponding to $n$ observations.    
 We randomly partition the dataset $\{\mathbf{x}_{i}\}_{1\leq i\leq n}$ 
 into $K$  disjoint subsets of approximately equal sizes, the $t^{th}$   subset
being of size $n_{t}\geq 2$ and $\displaystyle \sum_{t=1}^{K}n_{t}=n$. 
 For every $t$, let $\displaystyle \{  \mathbf{x}_{i}^{(t)}\}_{1\leq i\leq n_{t}}$ be the $t^{th}$  \textit{validation subset},  and its complement $\displaystyle  \{  \widetilde{\mathbf{x}}_{i}^{(t)} \}_{1\leq i\leq n-n_{t}}$, the $t^{th}$ \textit{training subset}.
 For every $t$  and for every pair $(\alpha_f, \alpha_b)$  of threshold parameters let $ \widehat{\mathcal{A}}_1^{(t)}, \ldots, \widehat{\mathcal{A}}_{p}^{( t)} $ be the   estimated neighborhoods  given by  GSA  using  the $t^{th}$   training subset.   For every $j=1,\ldots,p$ let $ \widehat{\beta}_{ \widehat{\mathcal{A}}_j^{(t)}}$   be the estimated coefficient   of the  regression of the variable $X_j$ on the neighborhood  $\widehat{ \mathcal{A} }^{(t)}_j $. 

Consider now the $t^{th}$  validation subset. So, for every $j$, using $\widehat{\beta}_{\mathcal{A}_j^{(t)}}^{(t)}$, we obtain  the vector of predicted values $\widehat{\mathbf{X}}_j^{(t)} \left(  \alpha_f, \alpha_b \right)$. If  $\mathcal{A}_j^{(t)}=\emptyset$ we predict each observation of    $X_j$ by the sample mean of the observations in the $t^{th}$ dataset of this variable.

Then,  we define the  $K$--fold cross--validation  function as
\begin{equation*}
CV\left(  \alpha_f, \alpha_b \right)=\frac{1}{n}\sum_{t=1}^{K}\sum_{j=1}^{p_j}  \left\|  \mathbf{X}_j^{(t)} - \widehat{\mathbf{X}}_j^{(t)}\left(  \alpha_f, \alpha_b \right) \right\|^2  
\end{equation*}
where $\left\|\cdot \right\|$ the L2-norm or euclidean distance in $\mathbb{R}^p$.    Hence the $K$--fold cross--validation forward--backward thresholds   $\widehat{\alpha}_f$, $\widehat{ \alpha}_b$  is 
\begin{equation*} 
\left( \widehat{\alpha}_f, \widehat{ \alpha}_b   \right)=:\mathop{\rm argmin}_{\left(  \alpha_f, \alpha_b \right)\in \mathcal{H}}CV\left(  \alpha_f, \alpha_b \right)
\end{equation*} 
where $\mathcal{H}$ is a grid of  ordered pairs $\left(  \alpha_f, \alpha_b \right)$  in $[0,1]\times [0,1]$ over which we perform
the search.
For a detail description see the Appendix.

\subsection {Example} \label{algor_example} 

\begin{figure} [H]
\begin{multicols}{3}
    \includegraphics[width=\linewidth]{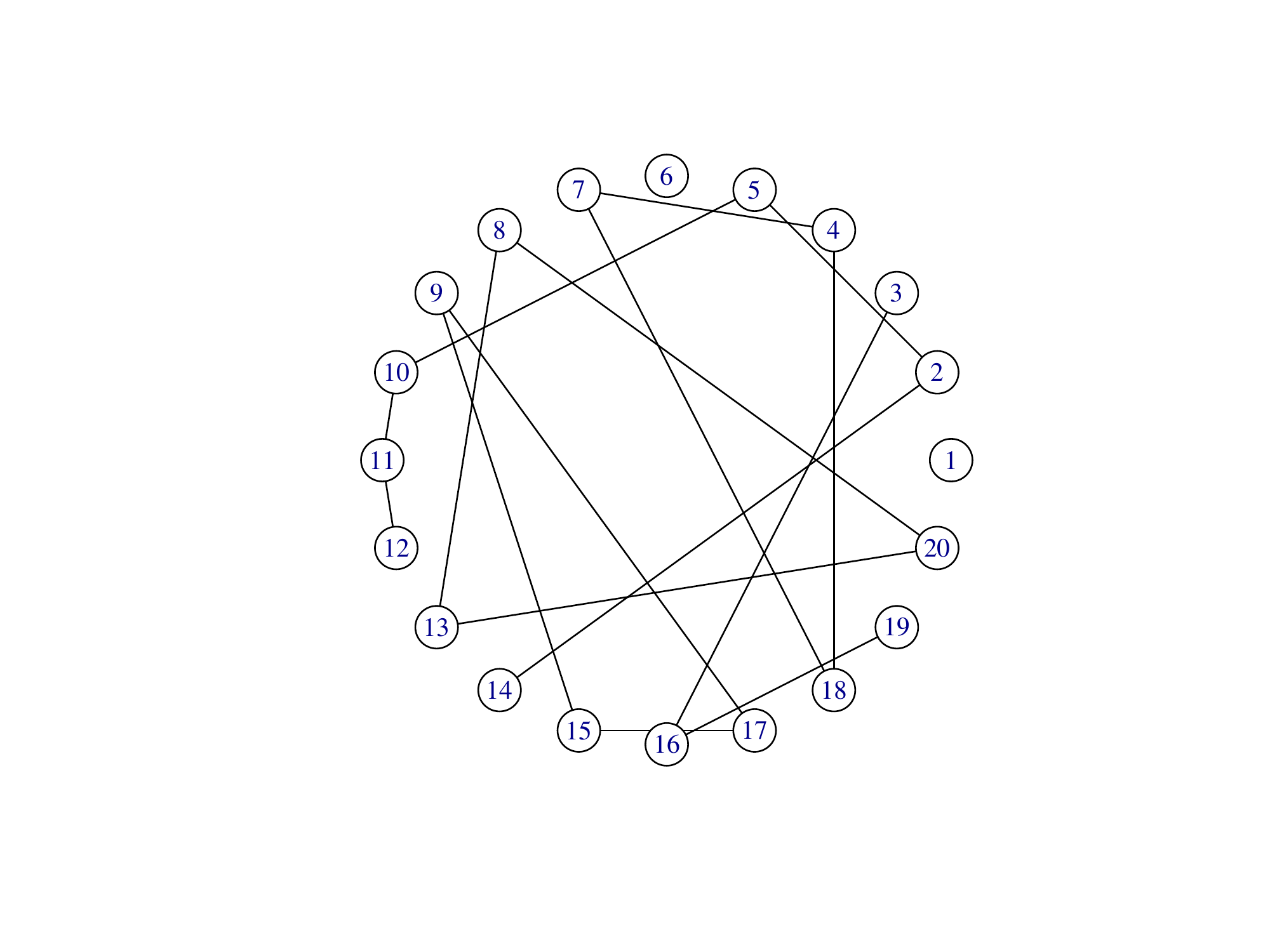}\par 
	\caption*{True graph}
    \includegraphics[width=\linewidth]{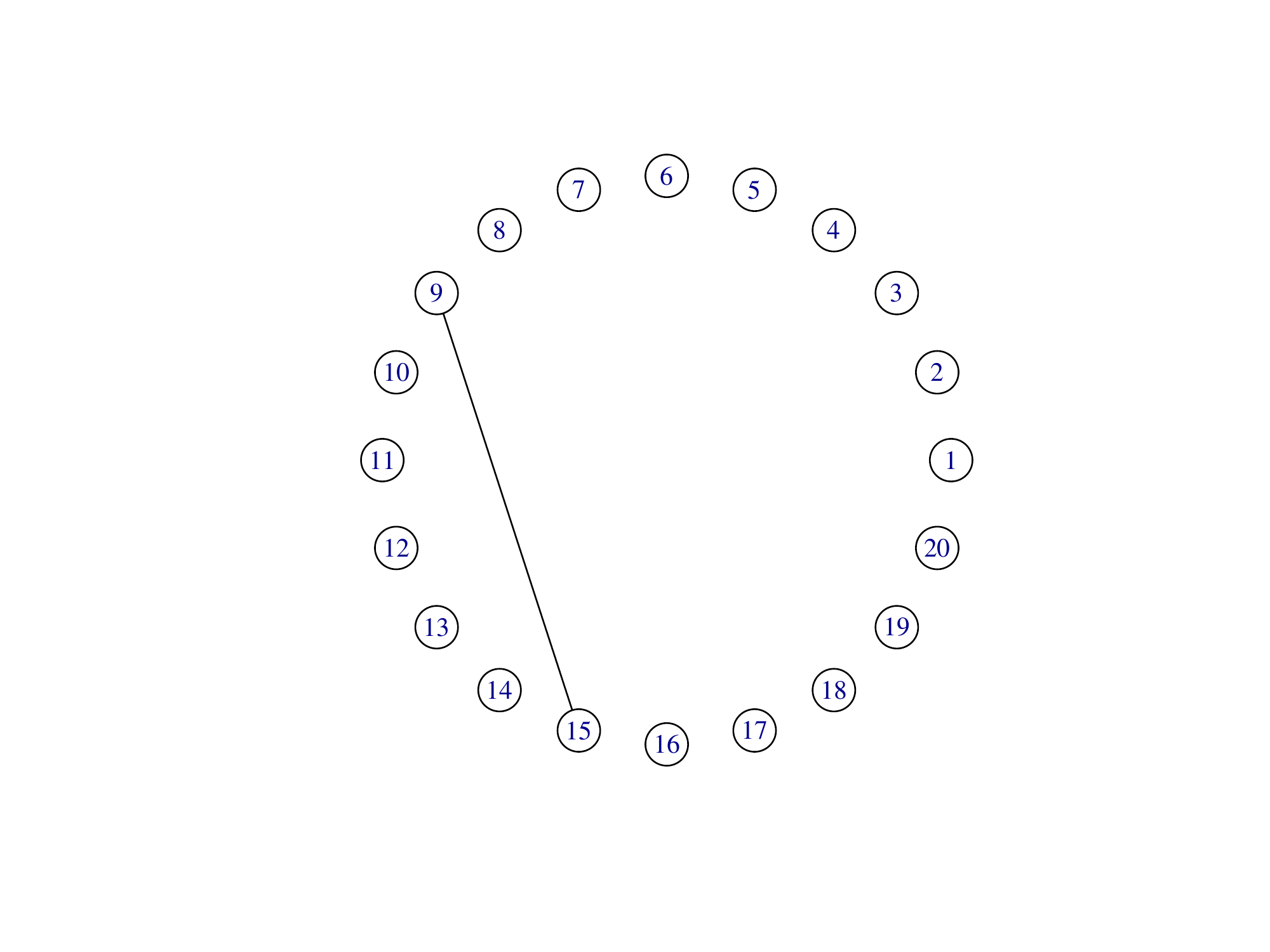}\par 
		\caption*{$k=1$}
		\includegraphics[width=\linewidth]{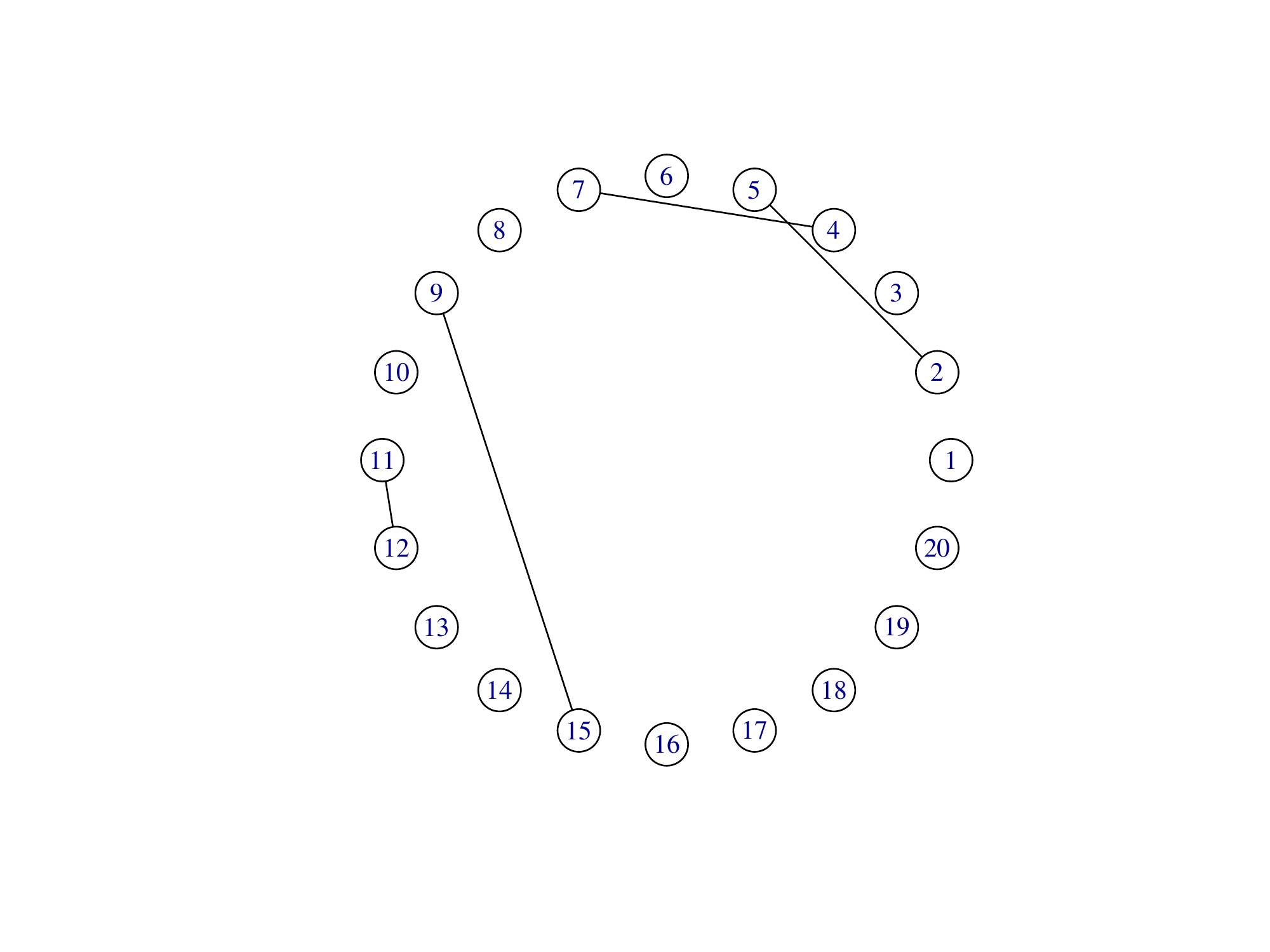}\par 
		\caption*{$k=4$}
    \end{multicols}
		\begin{multicols}{3}
    \includegraphics[width=\linewidth]{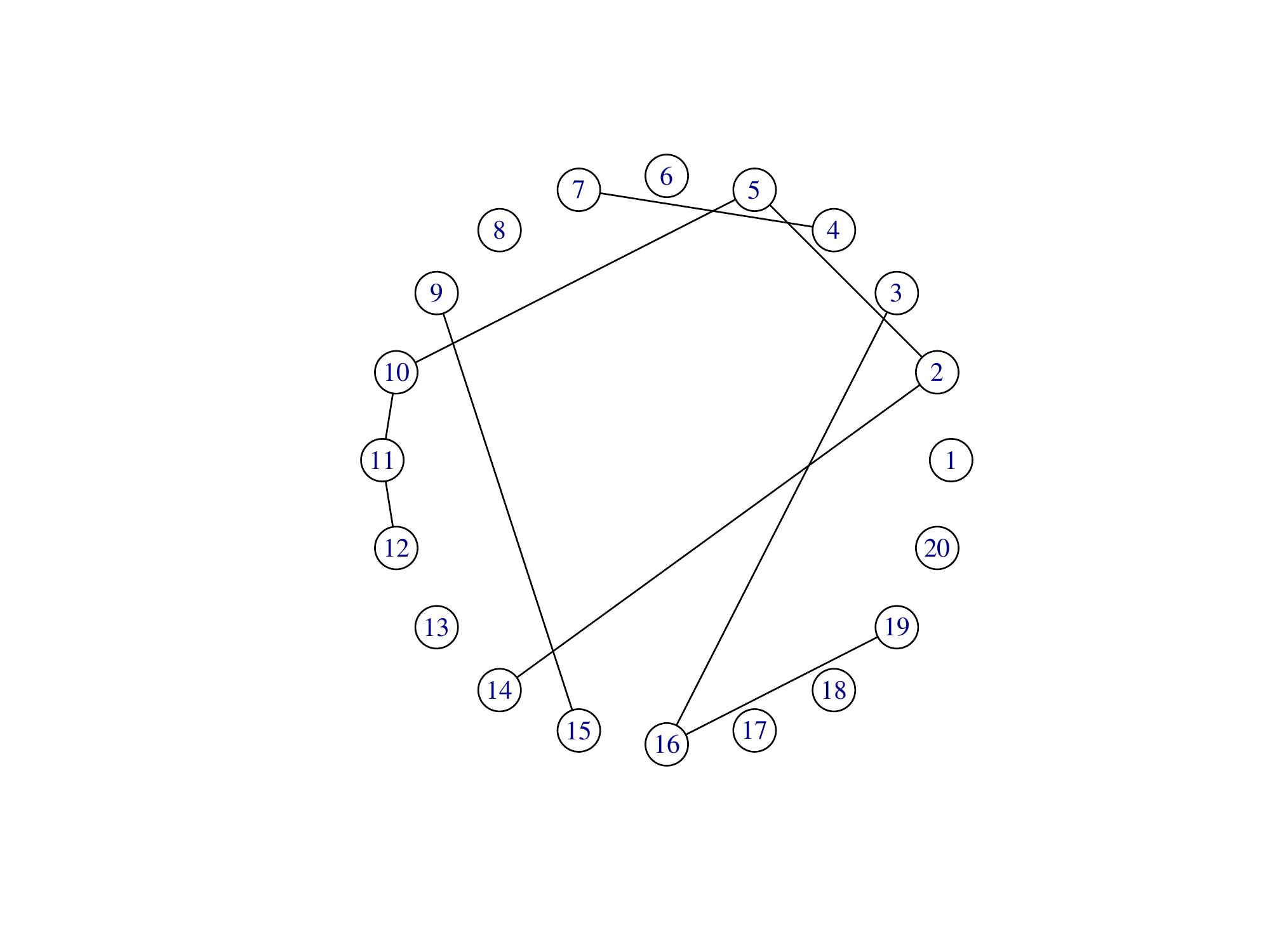}\par
			\caption*{$k=9$}
    \includegraphics[width=\linewidth]{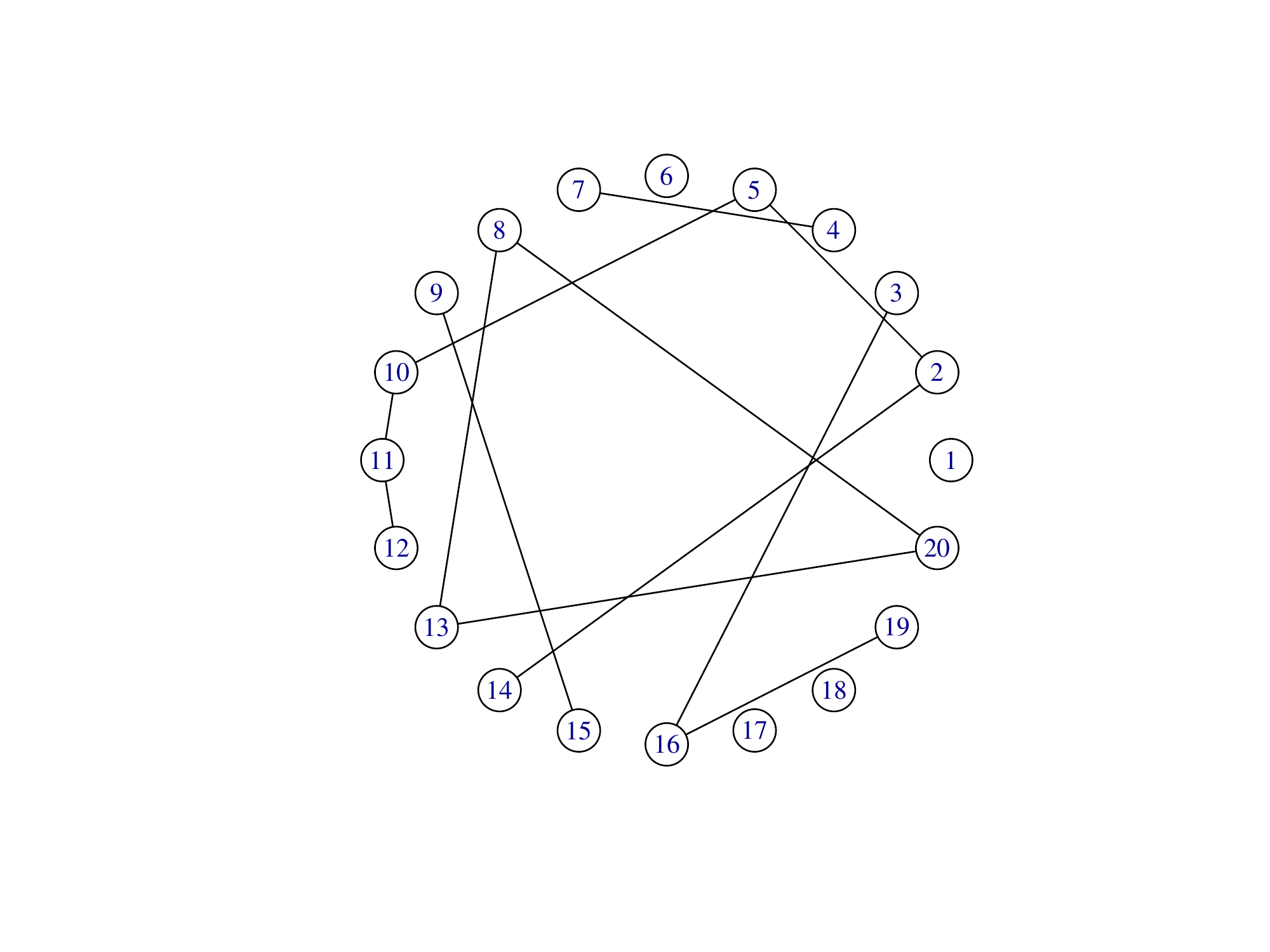}\par 
			\caption*{$k=12$}
		\includegraphics[width=\linewidth]{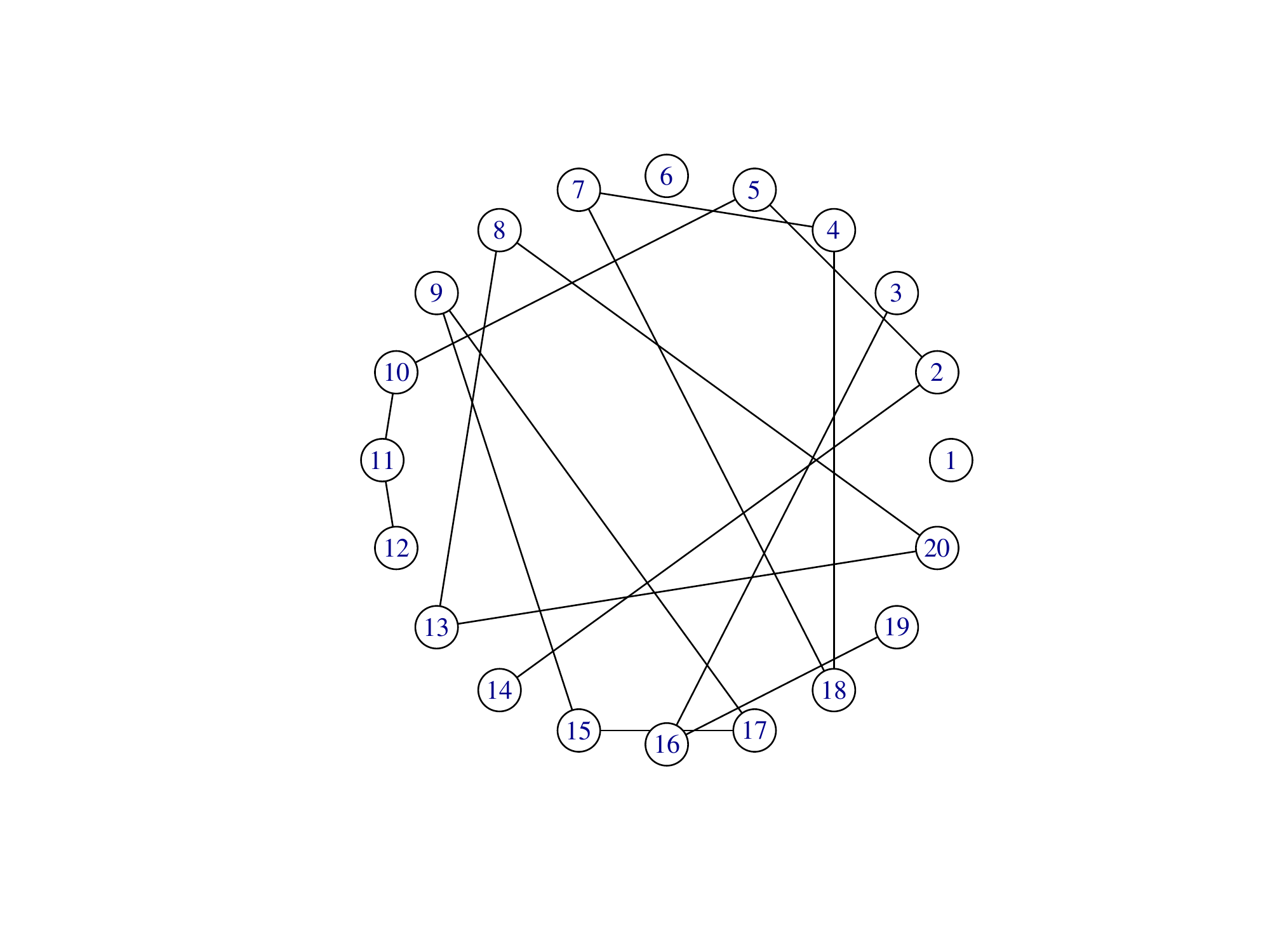}\par 
			\caption*{$k=16$}
    \end{multicols}
\caption{ True graph and sequence of successive  updates of  $\widehat{\mathcal{A}}_{j}^{k}$, for $k=1,4, 9, 12, 16$  of the GSA.}
\label{GGM_Gstepwise_Example}
\end{figure}

To illustrate the algorithm we consider  the GGM with 16  edges given  in  the first panel  of Figure  \ref{GGM_Gstepwise_Example}.  We draw $n=1000$ independent observations from this model (see the next section for details). The values for the threshold parameters $\alpha_f=0.17$ and $\alpha_b=0.09$ are determined by  $5$-fold cross-validation.  The figure also displays the selected pairs of edges at each step in a sequence of successive  updates of $\widehat{\mathcal{A}}_{j}^{k}$, for $k=1,4, 9, 12$ and the final step $k=16$, showing that the estimated graph is identical to the true graph.

\section{Numerical results and real data example }\label{section_3}
We conducted extensive Monte Carlo simulations to investigate  the performance of   GS.  In this section we report some results from this  study  and a numerical  experiment  using real data.

\subsection{Monte Carlo simulation study} \label{section_3.1}

 \noindent \textbf{Simulated Models}
	
\begin{figure} [H]
\begin{multicols}{3}
    \includegraphics[width=\linewidth]{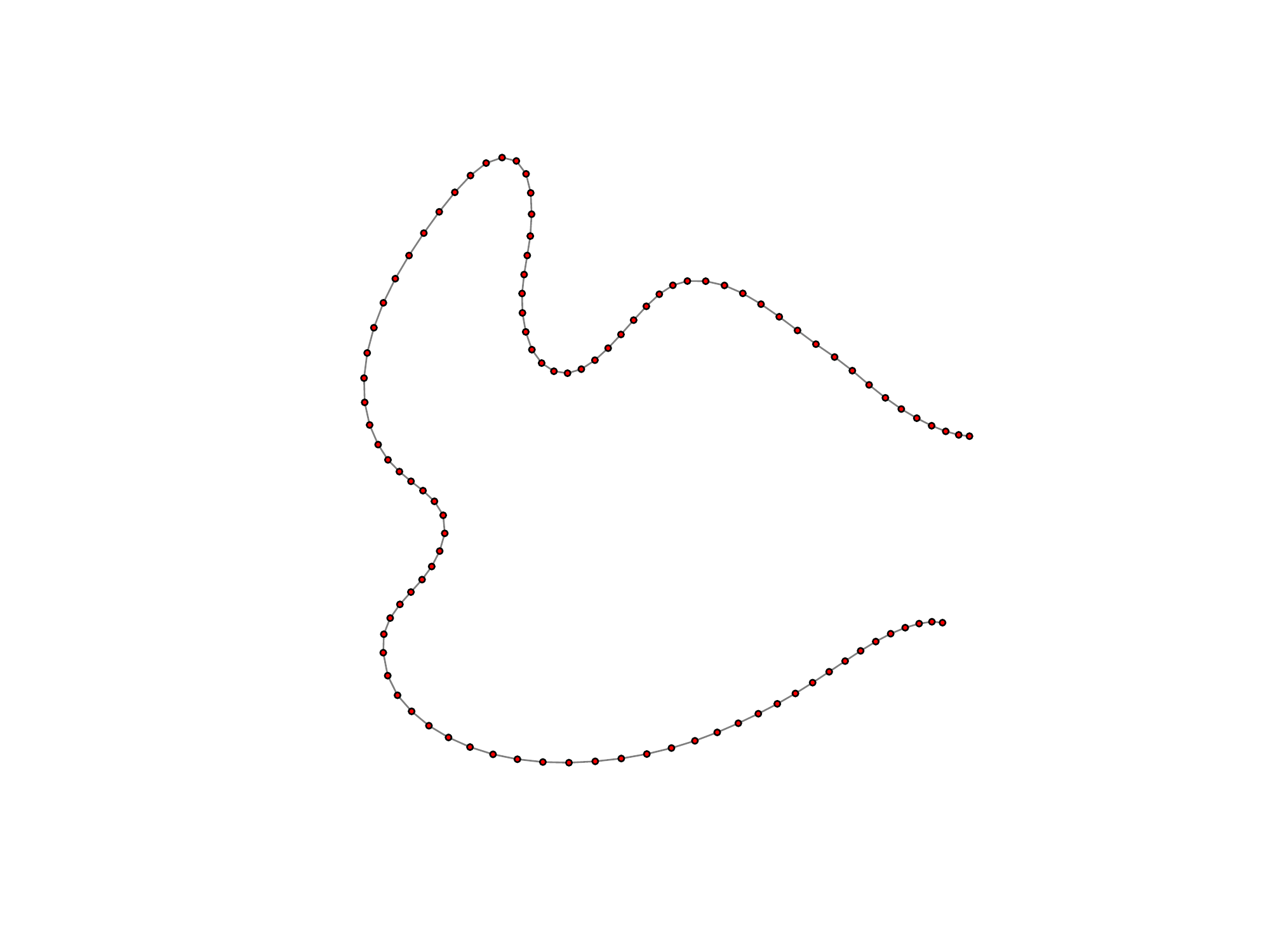}\par 
	\caption*{$\text{AR}(1)$}
    \includegraphics[width=\linewidth]{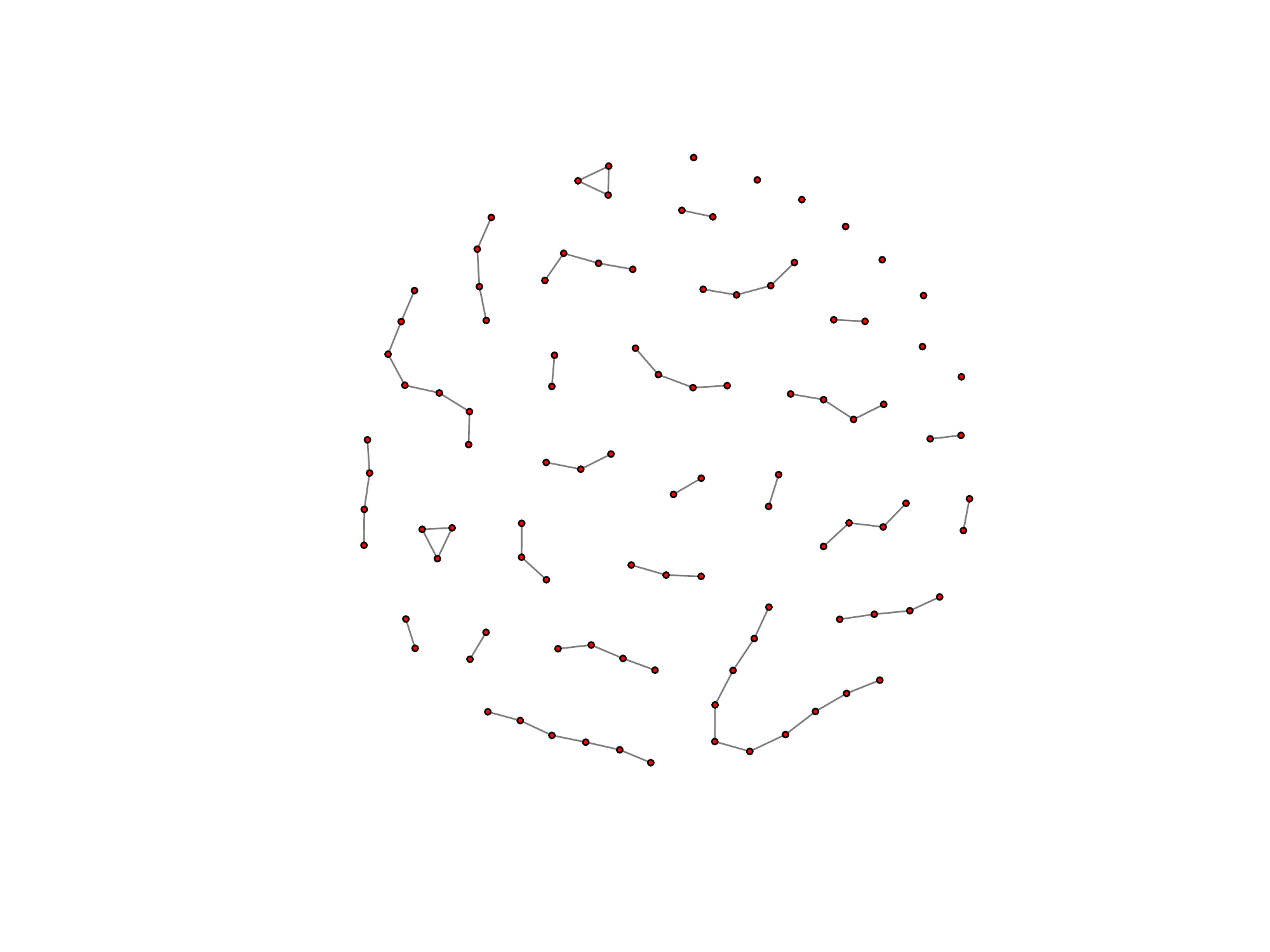}\par 
		\caption*{$\text{NN}(2)$}
		\includegraphics[width=\linewidth]{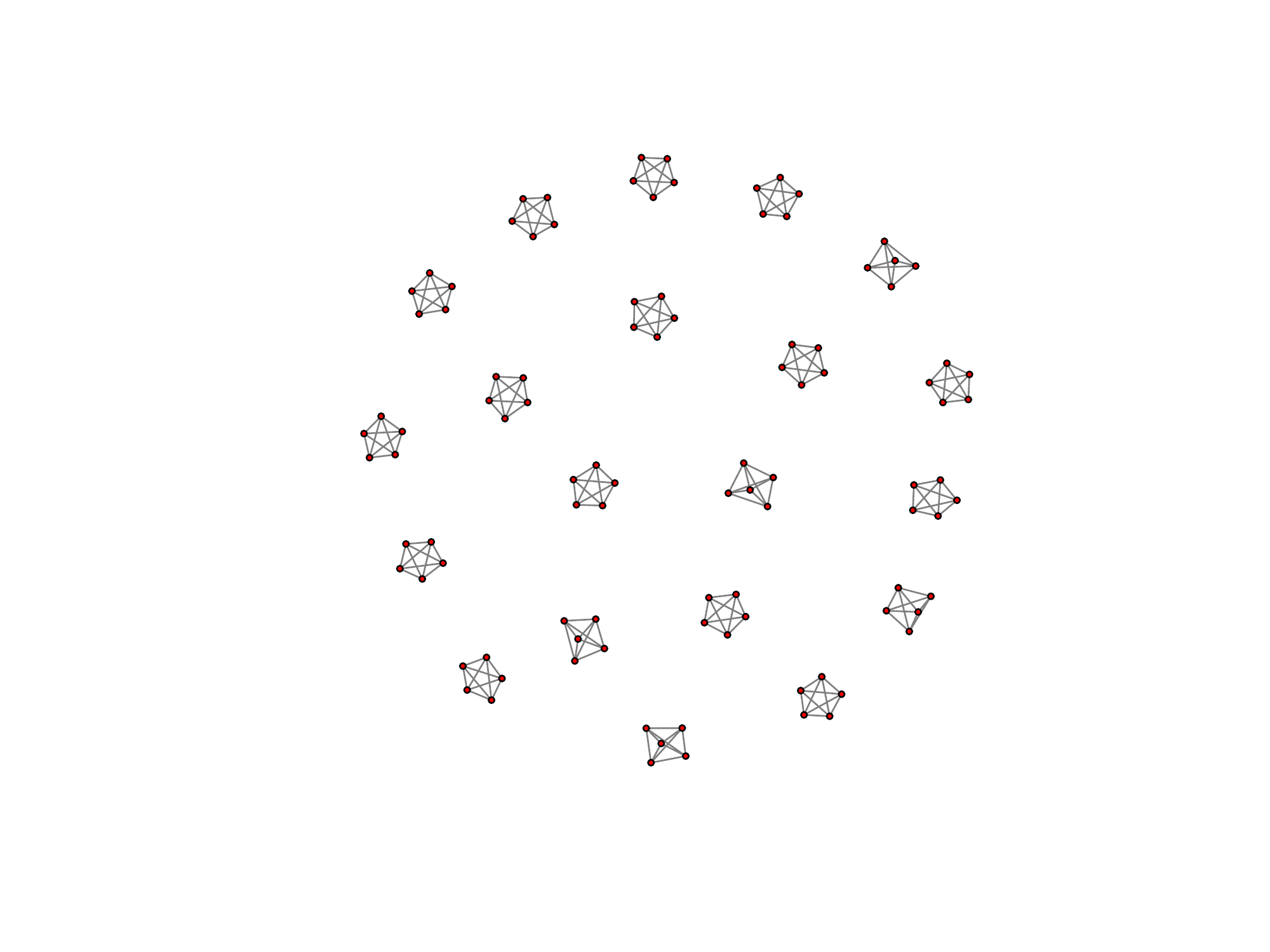}\par 
		\caption*{BG}
    \end{multicols}
\caption{ Graphs of $\text{AR}(1)$, $\text{NN}(2)$ and BG graphical models for $p=100$ nodes.}
\label{adjmatrices}
\end{figure}

 We consider three dimension values $p = 50, 100, 150$  and  three different models for  
 $\boldsymbol{\Omega}$:
\begin{itemize}
\item[]  {\bf Model 1.} Autoregressive model  of orden $1$, denoted $\text{AR}(1)$. In this case   $\Sigma_{ij}=0.4^{|i-j|}$ for $i,j=1,\ldots p$.   

\item[]  {\bf Model 2.}  Nearest neighbors model of order 2, denoted 
$\text{NN}(2)$. For each node  we randomly select two neighbors and choose a pair of symmetric entries of  $\boldsymbol{\Omega}$ using the NeighborOmega function of the R package Tlasso.    
\item[]  {\bf Model 3.}  Block diagonal matrix model with $q$ blocks of  size $p/q$, denoted BG. For $p=50,100$ and $150$,   we use $q=10, 20$ and $30$  blocks,  respectively. Each block, of size $p/q=5$,  has diagonal elements equal to $1$ and  off-diagonal elements equal to $0.5$.   
 \end{itemize}
 For each  $p$  and each model we generate $R=50$ random samples of size $n=100$. 
 These graph models are widely used  in the genetic literature to model gene expression data. See for example \cite{Lee} and  \cite{lighi}. Figure \ref{adjmatrices} displays graphs  from Models 1-3 with $p=100$ nodes. 

\noindent \textbf{Methods}

  We compare  the  performance of GS with Graphical lasso (Glasso) and Constrained $l_1$-minimization for inverse matrix estimation (CLIME) proposed by \cite{friedman2008sparse}  and \cite{cai2011constrained} respectively. Therefore, the methods compared in our simulation study are:
  
  \begin{itemize}
\item[{\bf 1.}]  The proposed method  GS with  the  forward and backward thresholds, $\left( {\alpha}_f, { \alpha}_b   \right)$,  estimated  by $5$-fold crossvalidation  on  a grid of $20 $ values   in   $[0,1]\times[0,1] $, as described in Subsection \ref{cross}.  The computing algorithm is available by request.
\item[{\bf 2.}] The Glasso  estimate  obtained by solving  the  $\ell_1$ penalized-likelihood problem: 
\begin{equation}\label{Glasso}
\min_{\boldsymbol{\Omega} \succ 0} \ \left( -\text{log}\{ \text{det}[\boldsymbol{\Omega}]\} + \text{tr}\{\boldsymbol{\Omega}\textbf{X}^{\top}\textbf{X}\} + \lambda \parallel \boldsymbol{\Omega} \parallel_1 \right).
\end{equation}
In our simulations and examples we use the R-package \textsc{CVglasso}     with the tuning parameter $\lambda$  selected by $5-$fold crossvalidation (the package default). 

\item[{\bf 3.}] The CLIME   estimate obtained by symmetrization of the solution of
\begin{equation}\label{Clime}
\min \{ \parallel \boldsymbol{\Omega} \parallel_1   \text{ subject to  } \left| S \boldsymbol{\Omega} -\boldsymbol{I}\right|_{\infty} \leq \lambda \},
\end{equation}
where $S$ is the sample  covariance, $\boldsymbol{I}$ is the identity matrix, $\left| \cdot \right|_{\infty}$ is the   elementwise $l_{\infty}$  norm, and $\lambda$ is a tuning parameter.  For computations, we use the R-package \textsc{clime} with the tuning parameter $\lambda$  selected by $5-$fold crossvalidation (the package default). 
\end{itemize}
To evaluate the ability of the methods for finding the pairs of edges, for each replicate,  we compute the  Matthews  correlation coefficient   \citep{matt}
\begin{equation}
\text{MCC}= \frac{\text{TP} \times \text{TN}   -    \text{FP} \times \text{FN}}{\sqrt{(\text{TP} + \text{FP})(\text{TP} + \text{FN})(\text{TN} + \text{FP})(\text{TN} + \text{FN})}},
\end{equation}
the $\text{Specificity} =  \text{TN}/(\text{TN}+\text{FP})$ and the $\text{Sensitivity} =  \text{TP}/(\text{TP}+\text{FN})$, where $\text{TP}$,  $\text{TN}$,  $\text{FP}$ and $\text{FN}$ are, in this order,  the number of true positives, true negatives,
false positives and false negatives, regarding the identification of  the nonzero off-diagonal elements of $\boldsymbol{\Omega}$. Larger values of MCC, Sensitivity and Specificity  indicate a better performance \citep{fan2009network, baldi2000}.    

For every replicate, the performance of 
$\widehat{\boldsymbol{\Omega}}$ as an estimate  for ${\boldsymbol{\Omega}}$  is measured by $m_F=|| \widehat{\boldsymbol{\Omega}}-\boldsymbol{\Omega} ||_{F}$  (where $|| \cdot||_{F}$  denotes the Frobenius norm) and by the  normalized Kullback-Leibler divergence defined by  $m_{NKL}=D_{KL}/(1+D_{KL})$ where 
\begin{eqnarray*}
D_{KL} =\frac12 \left(\text{tr} \left\{\widehat{\boldsymbol{\Omega}}\boldsymbol{\Omega}^{-1}\right\} - \text{log}\left\{\text{det}\left[\widehat{\boldsymbol{\Omega}}\boldsymbol{\Omega}^{-1}\right]\right\}-p\right)
\end{eqnarray*}
is the the Kullback-Leibler divergence between  $\widehat{\boldsymbol{\Omega}}$ and $\boldsymbol{\Omega}$.  
\medskip

\noindent \textbf{Results}

Table \ref{clasperf} shows the MCC performance for the three methods under   Models 1-3. GS clearly outperforms the other two methods while  CLIME just slightly outperforms  Glasso.    \cite{cai2011constrained} pointed out that  a procedure yielding a  more sparse 
$\widehat{\boldsymbol{\Omega}}$ is preferable because this  facilitates interpretation of the data.  The sensitivity and specificity results, reported in  Table \ref{clasperfapp} in  Appendix, show that  in general GS is more  sparse than the CLIME and Glasso, yielding fewer false positives (more  specificity) but  a few more false negatives (less sensitivity).       
Table \ref{Numerical_Performance} shows that under models $\text{AR}(1)$ and $\text{NN}(2)$ the three methods achieve fairly similar 
performances for estimating $\Omega$. However, under model BG,  GS clearly outperforms the other two.

\begin{figure}[H]
\centering
\subfloat[$p=50$]{\includegraphics[width = 0.8in]{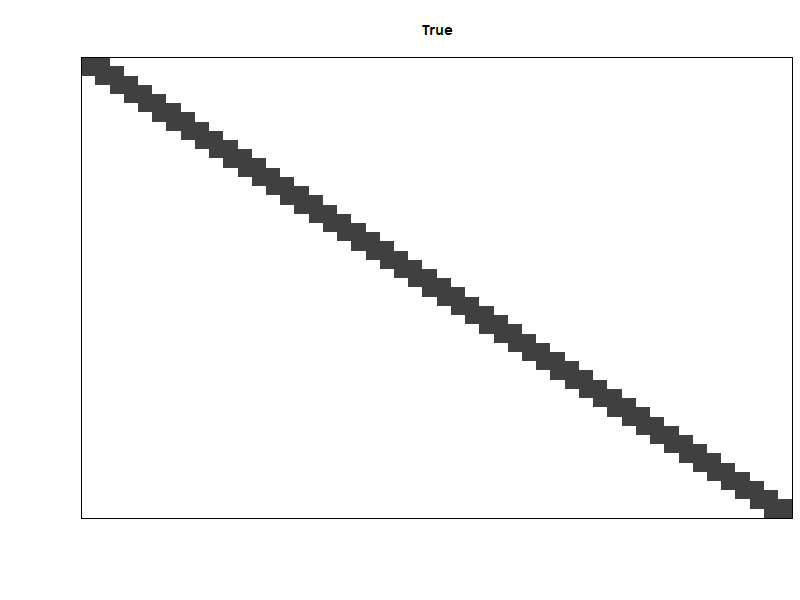}
\includegraphics[width = 0.8in]{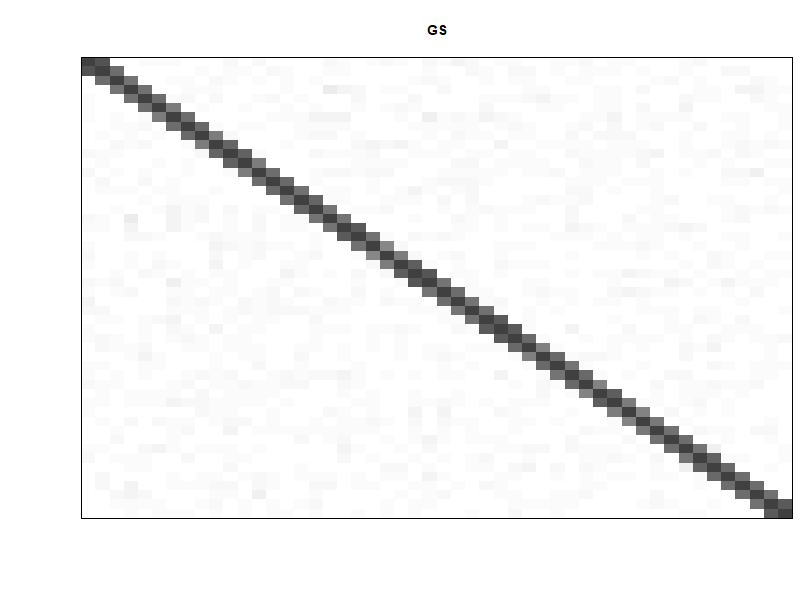}
\includegraphics[width = 0.8in]{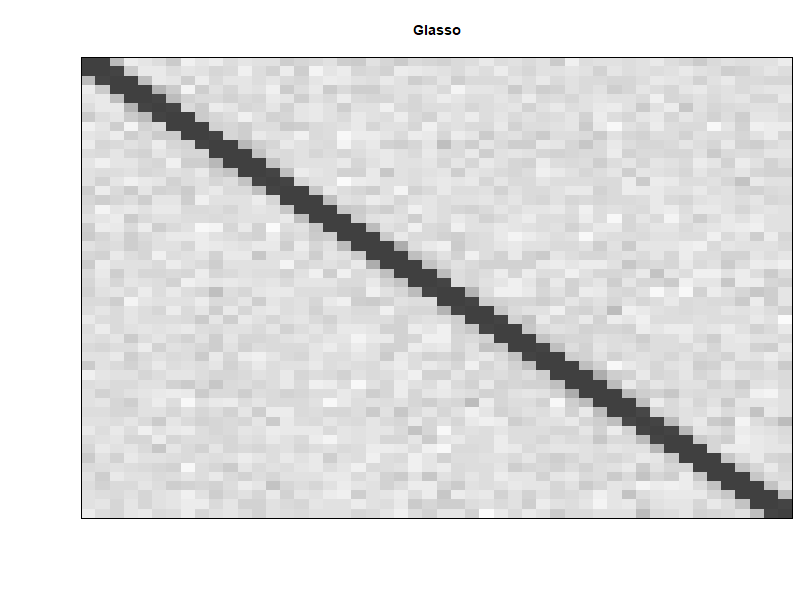}}
\includegraphics[width = 0.8in]{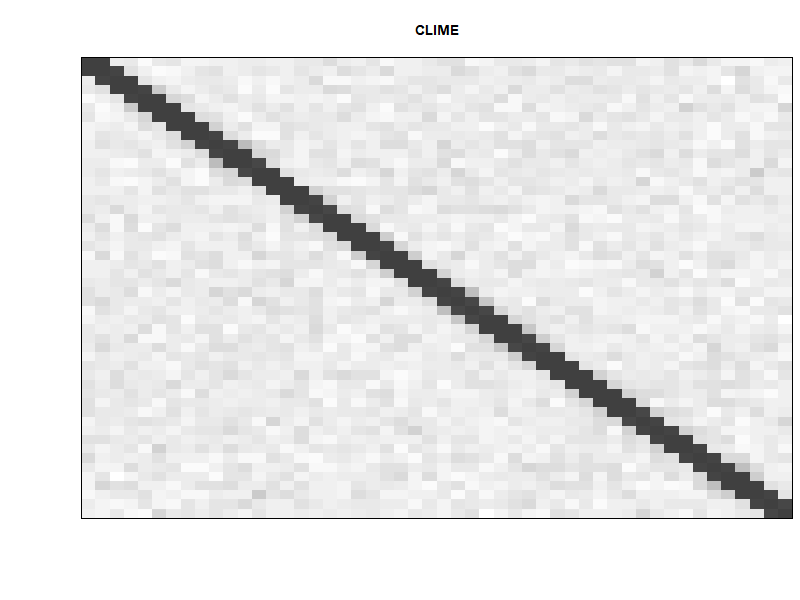} \\
\subfloat[$p=100$]{\includegraphics[width = 0.8in]{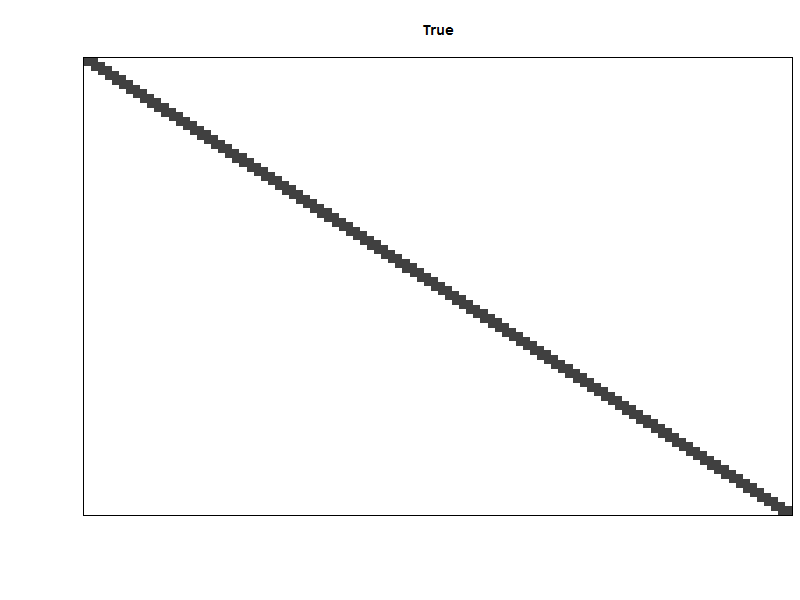}
\includegraphics[width = 0.8in]{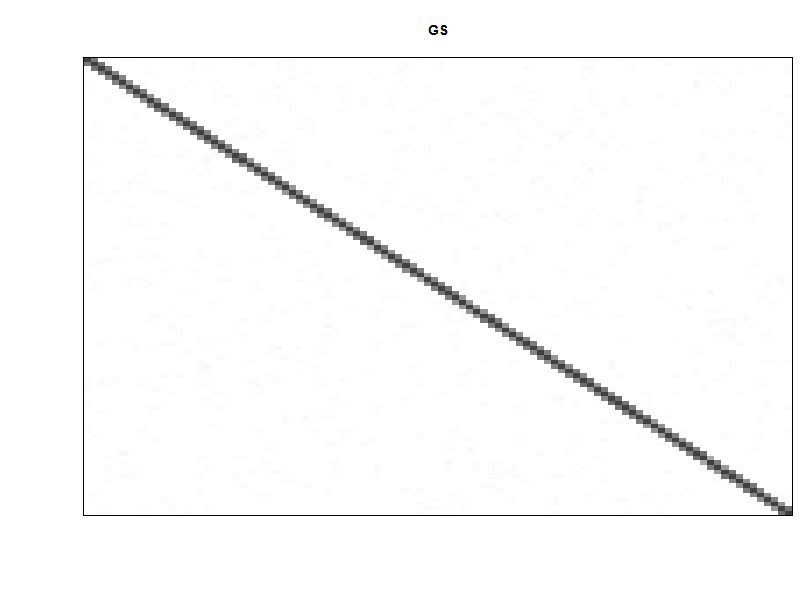}
\includegraphics[width = 0.8in]{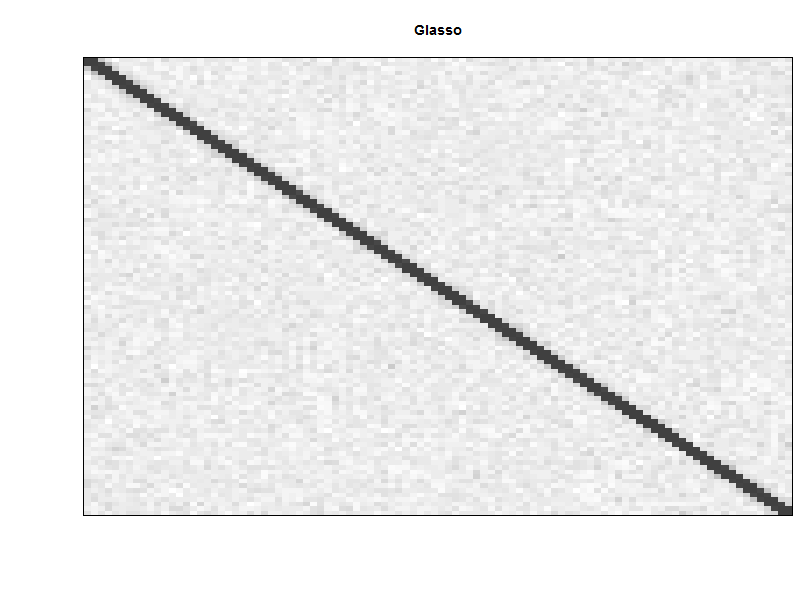}}
\includegraphics[width = 0.8in]{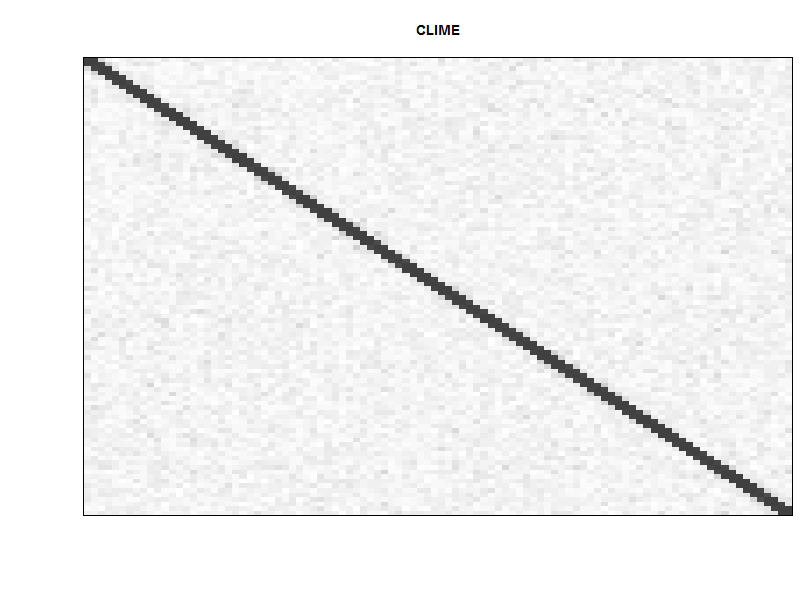} \\
\subfloat[$p=150$]{\includegraphics[width = 0.8in]{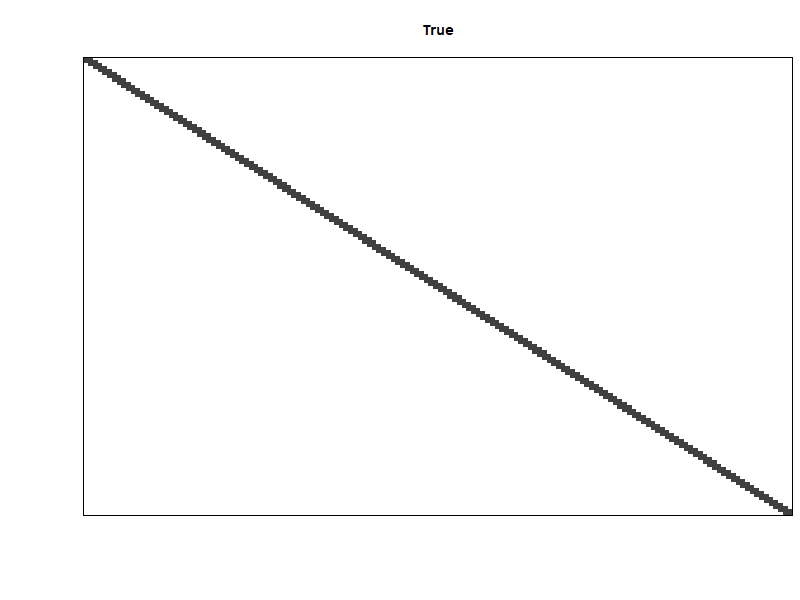}
\includegraphics[width = 0.8in]{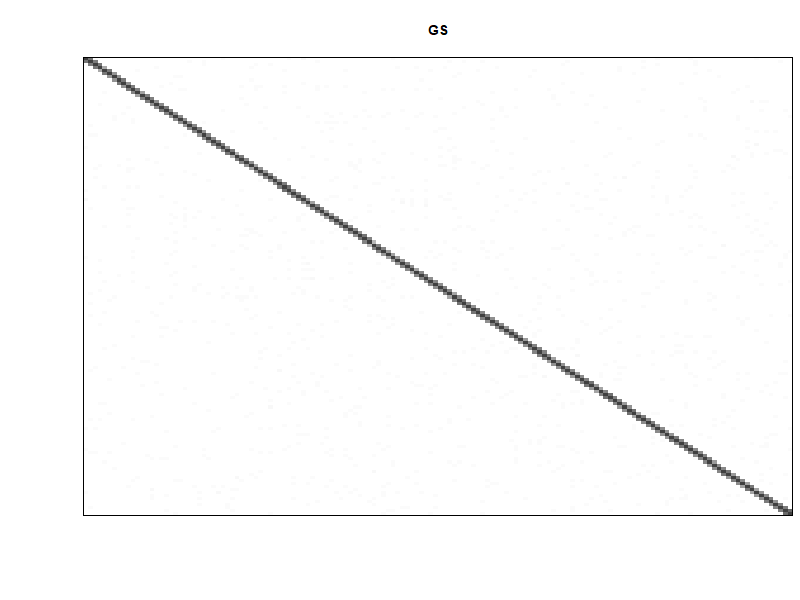}
\includegraphics[width = 0.8in]{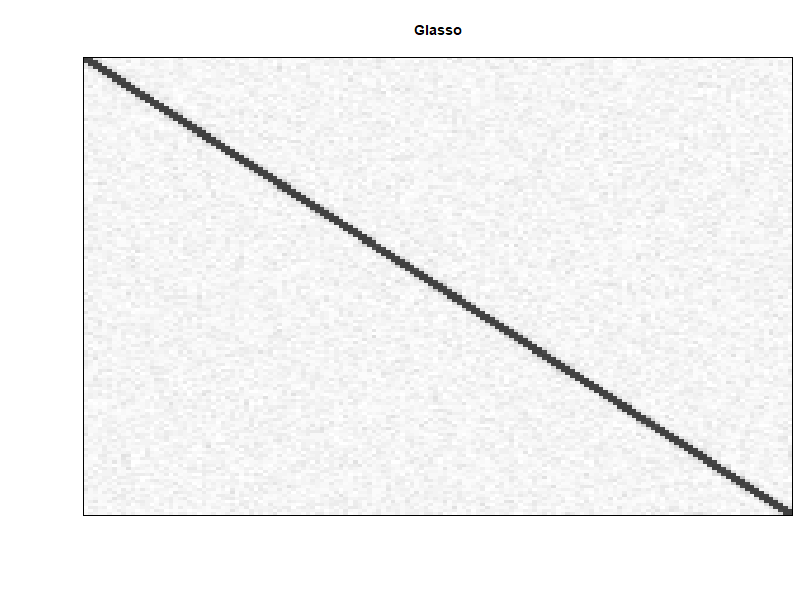}}
\includegraphics[width = 0.8in]{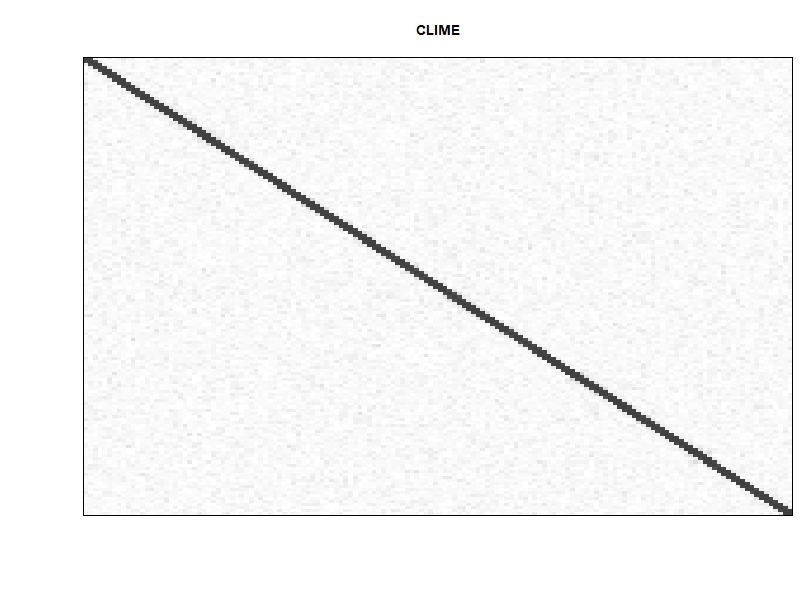} 
\caption{Model   $\text{AR}(1)$.  Heatmaps of the frequency of the zeros identified for each entry of the
precision matrix out of $R=50$ replicates. White color is 50 zeros
identified out of 50 runs, and black is 0/50.}
\label{Heatmaps_1}
\end{figure}

Figure \ref{Heatmaps_1}  display the heat-maps of the number of non-zero links identified in  the  $50$ replications  under model $\text{AR}(1)$. Notice that among the three compared methods,  the  GS  sparsity patterns best match those of  the true model.   Figures \ref{Heatmaps_2} and  \ref{Heatmaps_3}  in    the Appendix lead to similar conclusions for  models  $\text{NN}(2)$ and BG.

\begin{table}[H] 
\centering
\scriptsize
\caption{   Comparison  of  means and standard deviations (in brackets) of MCC over $R=50$ replicates.} 
\label{clasperf} 
\begin{tabular}{lrr|r|r}
\hline
    Model     &  $p$   & \multicolumn{1}{c|}{GS}         & \multicolumn{1}{c|}{Glasso}                   & \multicolumn{1}{c}{CLIME}                                                    \\ \hline
         & 50                      & 0.741   (0.009)                                &  0.419  (0.016)                 &    0.492 (0.006)              \\

$\text{AR}(1)$    & 100     & 0.751  (0.004)  & 0.433   (0.020) & 0.464   (0.004)         \\
                   
         & 150                 & 0.730  (0.004)  & 0.474   (0.017) & 0.499   (0.003)         \\ \hline

  &  50             & 0.751   (0.004)            &0.404   (0.014)              &   0.401 (0.007)            \\

$\text{NN}(2)$ & 100                  & 0.802  (0.005)  & 0.382  (0.006)                 & 0.407  (0.005)                  \\
       
  & 150            & 0.695   (0.007)            & 0.337   (0.008)              &   0.425 (0.003)            \\
			 \hline

      & 50           & 0.898  (0.005)            &  0.356 (0.009)     & 0.482   (0.005)          \\
         
BG    & 100     &   0.857  (0.005)     &  0.348   (0.004)   &   0.461  (0.002)  \\
   & 150       &   0.780  (0.008)     &  0.314   (0.003)   &   0.408  (0.003)  \\                                     
																					 \hline
																			 
\end{tabular}

\end{table}

\begin{table}[H]
\centering
\scriptsize
\caption{   Comparison  of  means and standard deviations (in brackets) of  $\text{m}_F$  and $\text{m}_{NKL}$ over $R=50$ replicates.} 
\label{Numerical_Performance}

\begin{tabular}{lrrr|rr|rr}
\hline 
 &   & \multicolumn{2}{c|}{GS}    & \multicolumn{2}{c|}{Glasso}  & \multicolumn{2}{c}{CLIME}
									\\ \hline
Model    & $p$                  & \multicolumn{1}{c}{$\text{m}_{NKL}$} & \multicolumn{1}{c|}{$\text{m}_F$ } &  \multicolumn{1}{c}{$\text{m}_{NKL}$} & \multicolumn{1}{c|}{$\text{m}_F$ }  & \multicolumn{1}{c}{$\text{m}_{NKL}$} & \multicolumn{1}{c}{$\text{m}_F$ }    \\ \hline
         & 50           & 0.70                   & 3.82                 & 0.64             & 3.90                  & 0.63      &   3.91          \\    
				 & \multicolumn{1}{l}{} & (0.00)      & (0.00)   & (0.00)  & ( 0.02)   & (0.00)  & (0.01)      \\
$\text{AR}(1)$   

 & 100                  & 0.83                  & 5.73           & 0.80             & 5.72  & 0.79          & 5.75                                          \\
         & \multicolumn{1}{l}{} & (0.00)                 & (0.00)                & (0.00)      & (0.02)   & (0.00)  & (0.01)  \\
        
				 & 150     & 1.25          & 7.16              &  1.17        & 7.21                 & 1.17        & 7.25       \\
         & \multicolumn{1}{l}{} & (0.00)                 & (0.00)           & (0.00)           & (0.02)                 & (0.00)                                         & (0.01)  \\
				
				 \hline

		  & 50    &  0.99                &  6.98         &   0.99                &   6.65       &   0.99         &6.64                                                                                      \\
         & \multicolumn{1}{l}{} & (0.00)     & (0.00)      &  (0.00)         & (0.01)         & (0.00)        & (0.00)                                                                                                   \\
$\text{NN}(2)$  & 100    &               0.10                 & 10.11                                 & 1.00                   & 9.64            &   1.00                                     & 9.601    
 \\
         & \multicolumn{1}{l}{} & (0.00)                 & (0.00)                                         & (0.00)                                         & (0.009)                 & (0.000)                                         & (0.005)                                         \\
             
						& 150    & 1.00          & 12.37               & 1.00       & 11.90                   & 1.00         & 11.79         \\
         & \multicolumn{1}{l}{} & (0.00)                 & (0.00)      & (0.00)             & (0.01)                 & (0.00)                 & (0.00)                                        
																			\\ \hline                                  
         
BG               & 50    & 0.46          & 1.44               & 0.85         &  5.45           & 0.82         & 5.03         \\
         & \multicolumn{1}{l}{} & (0.00)       & (0.00)          & (0.00)       & (0.10)         & (0.00)             & (0.05) \\

                & 100    & 0.71       & 2.94                & 0.93         & 9.16     & 0.92         & 8.71         \\
         & \multicolumn{1}{l}{} & (0.00)    & (0.00)      & (0.00)    & (0.07)                 & (0.00)                  & (0.02) \\

              & 150    & 0.88          &  6.10               &  0.96      & 11.59                   & 0.96          & 11.42         \\
         & \multicolumn{1}{l}{} & (0.00)                 & (0.00)                            & (0.00)            & (0.06)                 & (0.00)                                         & (0.02) \\                                
					\hline
\end{tabular}
 
\end{table}

\subsection{Analysis of Breast Cancer Data}\label{subsection_5_2}

 In preoperative chemoterapy, the complete eradication of all invasive cancer cells is referred to as {\it pathological complete response}, abbreviated as pCR.  It is  known in medicine that pCR is associated with  the long-term cancer-free survival of a patient.  Gene expression profiling (GEP) -- the measurement of the activity (expression level)  of genes in a patient -- could in principle be a useful predictor for  the patient's pCR.  

Using normalized gene expression data of patients in stages I-III of breast cancer,   \cite{hess2006pharmacogenomic} aim to identify patients that may achieve pCR  under {\em sequential anthracycline paclitaxel} preoperative chemotherapy.  When  a patient does not achieve  pCR state, he is classified in the group of  residual disease (RD), indicating that cancer still remains.    
Their data consist of 22283 gene expression levels for 133 patients, with 34 pCR and 99 RD. Following \cite{fan2009network} and \cite{cai2011constrained} we randomly  split the data into a  training set  and a testing set. The testing set is formed by randomly selecting 5 pCR patients and 16 RD patients (roughly  $1/6$ of the subjects) and the remaining patients form the training set. From the training set, a two sample t-test is performed to select the 50 most significant genes. The data is then standardized using the standard deviation estimated from the training set.

We apply a linear discriminant analysis (LDA) to predict whether a patient may achieve pathological complete response (pCR), based on the estimated inverse covariance
matrix of the gene expression levels.  We label with  $r=1$ the pCR group and $r=2$  the RD group and  assume that data are normally distributed,  with  common  covariance matrix $\boldsymbol{\Sigma}$ and different means $\boldsymbol{\mu}_r$.   From the training set, we obtain  $\widehat{\boldsymbol{\mu}}_r$, $\widehat{\boldsymbol{\Omega}}$ and for the test data compute the linear discriminant score as follows
\begin{equation}\label{LDA-score}
\delta_r(\textbf{x}) = \textbf{x}^{\top}\widehat{\boldsymbol{\Omega}} \widehat{\boldsymbol{\mu}}_r - \frac{1}{2} \boldsymbol{\mu}_r^{\top}\widehat{\boldsymbol{\Omega}} \boldsymbol{\mu}_r + \text{log} \widehat{\pi}_r \quad \text{for}\; i=1,\ldots,n,   
\end{equation}
where $\widehat{\pi}_r$ is the proportion of group $r$ subjects in the training set. The classification rule is
\begin{equation}\label{LD-rule}
\widehat{r}(\textbf{x}) = \argmax \delta_r (\textbf{x}) \;\;\; \text{for} \; r=1,2. 
\end{equation}
For every method  we use 5-fold cross validation on the training data to select the tuning constants. We repeat this scheme 100 times. 

 Table \ref{Breast_Cancer_Table} displays the means and standard errors (in brackets) of Sensitivity, Specificity, MCC and Number of selected  Edges using $\widehat{\boldsymbol{\Omega}}$ over the 100 replications.  Considering the  MCC, GS is slightly better than CLIME and CLIME than Glasso. While the three methods give similar  performance considering the Specificity,  GS and CLIME improve over Glasso in terms of Sensitivity.

\begin{table}[H]
\centering
\scriptsize
  
\caption{ Comparison  of  means and standard deviations (in	 brackets) of Sensitivity, Specificity,  MCC and Number of selected edges over 100 replications.}
\label{Breast_Cancer_Table}
\begin{tabular}{lrrrr}
\hline
                   & \multicolumn{1}{c}{GS } & \multicolumn{1}{c}{CLIME} & \multicolumn{1}{c}{Glasso} & \multicolumn{1}{l}{} 
									\\ \hline
Sensitivity        & 0.798  (0.02)                  & 0.786  (0.02)                  & 0.602  (0.02)                                      \\
                   
Specificity        & 0.784 (0.01)                   & 0.788  (0.01)                  & 0.767 (0.01)                                                 \\
                   
MCC                & 0.520 (0.02)                   & 0.516  (0.02)                  &   0.334 (0.02)                                                   \\
                   
Number of Edges    & 54 (2)                         & 4823 (8)                       &  2103  (76)                                                                                        \\

									\hline
\end{tabular}
\end{table}

\section{Concluding remarks}\label{section_4}

This paper introduces a stepwise  procedure, called  GS, to perform covariance selection  in high dimensional Gaussian graphical models. Our method uses a different parametrization of the Gaussian graphical model based on Pearson correlations  between the best-linear-predictors prediction errors.   The GS  algorithm  begins with a family of empty neighborhoods and using basic steps,  forward and  backward, adds or delete edges until  appropriate thresholds for  each step are reached.   These thresholds are automatically determined  by cross--validation. 

GS is compared with Glasso and CLIME under different Gaussian graphical models ($\text{AR}(1)$, $\text{NN}(2)$ and BG) 
and using different performance measures regarding network recovery and sparse estimation of the precision matrix  $\Omega$. 
GS is shown to have good support recovery performance and to produce  simpler models than  the other two methods (i.e. GS is a parsimonious estimation procedure). 

We use GS for the analysis of  breast cancer data and show  that this method may be a useful tool for applications in medicine and other fields.

\section*{Acknowledgements}

The authors thanks the generous support of NSERC, Canada, the Institute of Financial Big Data, University Carlos III of Madrid and the CSIC, Spain.

\appendix

\section{ Appendix }

\subsection{Selection of the thresholds parameters by cross-validation}

Let   $\boldsymbol{X}$ be the   $n\times p$ matrix with  rows $\mathbf{x}_{i}=\left(x_{i1},\ldots,x_{ip} \right)$, $i=1,\ldots,n$,  corresponding to $n$ observations.
For each $j=1,\ldots, p$, let $\mathbf{X}_j=\left( x_{1j}, \ldots, x_{nj}  \right)^{\top}$ denote the jth--column of the matrix $\mathbf{X}$.

 We randomly partition the dataset $\{\mathbf{x}_{i}\}_{1\leq i\leq n}$ 
 into $K$  disjoint subsets of approximately equal size, the $t^{th}$   subset
being of size $n_{t}\geq 2$ and $\displaystyle \sum_{t=1}^{K}n_{t}=n$. 
 For every $t$, let $\displaystyle \{  \mathbf{x}_{i}^{(t)}\}_{1\leq i\leq n_{t}}$ be the $t^{th}$  \textit{validation subset},  and its complement $\displaystyle  \{  \widetilde{\mathbf{x}}_{i}^{(t)} \}_{1\leq i\leq n-n_{t}}$, the $t^{th}$ \textit{training subset}. 

For every $t=1,\ldots, K$  and  threshold parameters $(\alpha_f, \alpha_b)\in [0,1]\times [0,1]$  let $ \widehat{\mathcal{A}}_1^{(t)}, \ldots, \widehat{\mathcal{A}}_{p}^{(t)} $ be the  estimated neighborhoods  given by  GSA  using  the $t^{th}$   training subset $\displaystyle  \{  \widetilde{\mathbf{x}}_{i}^{(t)} \}_{1\leq i\leq n-n_{t}}$ with $   \widetilde{\mathbf{x}}_{i}^{(t)} = (\widetilde{x}_{i1}^{(t)},\ldots, \widetilde{x}_{ip}^{(t)}),$ $ 1\leq i \leq n-n_{t}$. Consider  for every node $j$ the estimated  neighborhood  $\widehat{ \mathcal{A} }^{(t)}_j=\left\{l_1,\ldots, l_q\right\}$      and let $ \widehat{\beta}_{ \widehat{\mathcal{A}}_j^{(t)}}$ be the estimated coefficient of the regression of $\widetilde{\mathbf{X}}_j=(\widetilde{x}_{1j}^{(t)},\ldots, \widetilde{x}_{n-n_tj}^{(t)})^{\top}$ on $X_{l_1},\ldots, X_{l_q}$, represented in \eqref{matrixfig} (red colour).

Consider  the $t^{th}$  validation subset $\displaystyle \{  \mathbf{x}_{i}^{(t)}\}_{1\leq i\leq n_{t}}$  with  $   \mathbf{x}_{i}^{(t)} =( x_{i1}^{(t)},\ldots, x_{ip}^{(t)})$, $ 1 \leq i \leq n_t $ and for every $j$ let  $\mathbf{X}_j^{(t)}=\left( x_{1j}^{(t)}, \ldots, x_{n_{t}j}^{(t)}  \right)^{\top}$  and define the vector of predicted values
\begin{eqnarray*}
  \widehat{\mathbf{X}}_j^{(t)} \left(  \alpha_f, \alpha_b \right) = \mathbf{X}_{ \widehat{\mathcal{A}}_j^{ (t) }} \widehat{\beta}_{ \mathcal{A}_j^{(t)}}^{(t)},
\end{eqnarray*}
where $\mathbf{X}_{ \widehat{\mathcal{A}}_j^{ (t) }}$ is the matrix with rows $( x_{il_1}^{(t)},\ldots,  x_{il_q}^{(t)}   )$, $ 1 \leq i \leq n_t $ represented in \eqref{matrixfig} (in blue colour).   If the neighborhood  $\mathcal{A}_j^{(t)}=\emptyset$ we define  
\begin{eqnarray*}\widehat{\mathbf{X}}_j^{(t)}\left(  \alpha_f, \alpha_b \right)=(\bar{x}_{j}^{(t)},\ldots, \bar{x}_{j}^{(t)})^{\top}
\end{eqnarray*}
where $\bar{x}_{j}^{(t)}$ is the mean of the sample of observations $x_{1j}^{(t)}, \ldots, x_{n_{t}j}^{(t)} $.

We define the  $K$--fold cross--validation  function   as
\begin{equation*}
CV\left(  \alpha_f, \alpha_b \right)=\frac{1}{n}\sum_{t=1}^{K}\sum_{j=1}^{p}  \left\|  \mathbf{X}_j^{(t)} - \widehat{\mathbf{X}}_j^{(t)}\left(  \alpha_f, \alpha_b \right) \right\|^2 
\end{equation*}
where $\left\|\cdot \right\|$ the L2-norm or euclidean distance in $\mathbb{R}^p$.   Hence the $K$--fold cross--validation forward--backward thresholds   $\widehat{\alpha}_f$, $\widehat{ \alpha}_b$  is
\begin{equation} \label{threscross}
\left( \widehat{\alpha}_f, \widehat{ \alpha}_b   \right)=:\mathop{\rm argmin}_{\left(  \alpha_f, \alpha_b \right)\in \mathcal{H}}CV\left(  \alpha_f, \alpha_b \right)
\end{equation} 
where $\mathcal{H}$ is a grid of  ordered pairs $\left(  \alpha_f, \alpha_b \right)$  in $[0,1]\times [0,1]$ over which we perform
the search.

\begin{small}
\begin{eqnarray} \label{matrixfig}
\left( \begin{array}{c  c ccccccccc}
 t^{th} & \text{training} &  \text{subset }    \\
\cdots &  \textcolor[rgb]{1,0,0}{\widetilde{x}_{1j}^{(t)}} &  \cdots & \textcolor[rgb]{1,0,0}{\widetilde{x}_{1l_1}^{(t)}} & \cdots & \textcolor[rgb]{1,0,0}{\widetilde{x}_{1l_q}^{(t)}} &   \cdots    \\
\vspace{0.5cm}
	\vdots & \vdots & \vdots & \vdots & \vdots&  \vdots &  \vdots\\
	\vspace{0.5cm}
	
	\cdots &  \textcolor[rgb]{1,0,0}{\widetilde{x}_{n-n_tj}^{(t)}} &  \cdots   & \textcolor[rgb]{1,0,0}{\widetilde{x}_{n-n_t l_1}^{(t)}} & \cdots & \textcolor[rgb]{1,0,0}{\widetilde{x}_{n-n_tl_q}^{(t)}} & \cdots \\
   \midrule
	t^{th}  &  \text{validation} &  \text{subset }     \\
	     \cdots &  \textcolor[rgb]{0,0,1}{x_{1j}^{(t)}} &  \cdots & \textcolor[rgb]{0,0,1}{x_{1l_1}^{(t)}} & \cdots & \textcolor[rgb]{0,0,1}{x_{1l_q}^{(t)}} &     \cdots    \\
		\vdots & \vdots & \vdots & \vdots & \vdots&  \vdots &  \vdots\\
	\cdots &  \textcolor[rgb]{0,0,1}{x_{n_tj}^{(t)}} &  \cdots   & \textcolor[rgb]{0,0,1}{x_{n_t l_1}^{(t)}} & \cdots & \textcolor[rgb]{0,0,1}{x_{n_tl_q}^{(t)}} &     \cdots  \\
\end{array}\right)
\end{eqnarray} 
\end{small}
\begin{remark}
Matrix \eqref{matrixfig} represents, for every node $j$ the comparison between estimated and predicted values for cross-validation.   $\widehat{\beta}_{ \widehat{\mathcal{A}}_j^{(t)}}$ is computed using the observations $\widetilde{\mathbf{X}}_j=(\widetilde{x}_{1j}^{(t)},\ldots, \widetilde{x}_{n-n_tj}^{(t)})^{\top}$ and the matrix $\widetilde{ \mathbf{X}}_{ \widehat{\mathcal{A}}^{(t)}_j } $   with rows $(\widetilde{x}_{il_1}^{(t)},\ldots, \widetilde{x}_{il_q}^{(t)})$, $i=1,\ldots,n-n_t$ in the $t^{th}$ training subset (red colour). Based on the $t^{th}$ validation set $\widehat{\mathbf{X}}_j^{(t)}$ is computed using $\mathbf{X}_{ \widehat{\mathcal{A}}_j^{ (t) }}$   and compared with   $\mathbf{X}_j$ (in blue color).
\end{remark}

\newpage

\subsection{ Complementary simulation results }

\begin{table}

\centering 
\scriptsize
 \caption{   Comparison  of  means  and standard deviations (in brackets) of Specificity, Sensitivity and MCC over $R=50$ replicates. } 
	\label{clasperfapp} 

\begin{tabular}{lrrrr|rrr|rrr}
\hline
         &     & \multicolumn{3}{c|}{GS}         & \multicolumn{3}{c|}{Glasso}                   & \multicolumn{3}{c}{CLIME}                                                    \\ \hline
Model    & $p$                  & \multicolumn{1}{c}{Sensitivity} & \multicolumn{1}{c}{Specificity} & \multicolumn{1}{c|}{MCC} & \multicolumn{1}{c}{Sensitivity} & \multicolumn{1}{c}{Specificity} & \multicolumn{1}{c|}{MCC} & \multicolumn{1}{c}{Sensitivity} & \multicolumn{1}{c}{Specificity} & \multicolumn{1}{c}{MCC} \\ \hline
         & 50 &  0.756                    & 0.988                       & 0.741                 & 0.994     & 0.823    &  0.419     & 0.988                                             &     0.891               &    0.492               \\
         & \multicolumn{1}{l}{} & (0.015)                  & (0.002)                  & (0.009)                 & (0.002)                  & (0.012)                  & (0.016)                 & (0.002)                  & (0.003)                  & (0.006)                 \\

$\text{AR}(1)$    & 100     & 0.632 & 0.999  & 0.751  & 0.989   & 0.897    & 0.433   & 0.983  & 0.934         & 0.464           \\
         & \multicolumn{1}{l}{} & (0.007)        & (0.000)                  & (0.004)                 & (0.002)                  & (0.009)                  & (0.020)                 & (0.002)                  & (0.001)                  & (0.004) \\
				
				& 150     & 0.607 & 0.999  & 0.730  & 0.981   & 0.943   & 0.474   & 0.972  & 0.964        & 0.499           \\
         & \multicolumn{1}{l}{} & (0.006)        & (0.000)                  & (0.004)                 & (0.002)                  & (0.007)                  & (0.017)                 & (0.002)                  & (0.001)                  & (0.003) \\ \hline

  &  50        & 0.632       & 0.999     & 0.751    & 0.971      & 0.864         &0.404          &  0.984       &   0.875      &   0.401             \\
& \multicolumn{1}{l}{} & (0.007)          & (0.000)                  & (0.004 )               & (0.004)                  & (0.010)    & (0.014)   & (0.003)                  & (0.004)                  & (0.007)                 \\
$\text{NN}(2)$ & 100                  & 0.730                    & 0.999                   & 0.802                 & 0.987              & 0.924            & 0.382                   & 0.985                    & 0.937                   & 0.407                  \\
         & \multicolumn{1}{l}{} & (0.008)           & (0.000)         & (0.005)         & (0.002)         & (0.004)            & (0.006)                 & (0.002)                  & (0.001)                  & (0.005)                 \\
         & 150       &  0.555            &  0.999                    & 0.695                   & 0.952                    & 0.936                   & 0.337                   &  0.934                    & 0.965                    & 0.425                   \\
        & \multicolumn{1}{l}{} & (0.017)                  & (0.000)                  & (0.007)                 & (0.004)                  & (0.002)                  & (0.008)                 & ( 0.003)                  & (0.001)                  & (0.003)       \\          
			 \hline

                 & 50    & 0.994          & 0.981      & 0.898     &  0.867   & 0.697         &  0.356  & 0.962    &  0.807     & 0.482             \\
         & \multicolumn{1}{l}{} & (0.002)      & (0.001)     & (0.005)    & (0.032)    & (0.021)    & (0.009)     & (0.004)                 & (0.005)                                                                          
                                       & (0.005)                                         \\ 
BG    & 100    &    0.949       &   0.989    & 0.857     &0.569      &0.908      &  0.348     & 0.818        &0.920  & 0.4615                  \\
         & \multicolumn{1}{l}{} & (0.007)                 & (0.000)    & (0.005)     & (0.039)  & (0.011)    & ( 0.004)       & (0.005)         & (0.005)      & (0.002)                                         \\ 
          & 150     & 0.782    & 0.994    &  0.780   & 0.426  &  0.952    & 0.314   & 0.626  &0.959      & 0.408          \\
         & \multicolumn{1}{l}{} & (0.021)     & (0.000)  & (0.008)     & (0.035)    & (0.006)   & (0.003) & (0.006)    & (0.001)  & (0.003)         \\
																				 \hline
																			 
\end{tabular}

\end{table}

\begin{figure}[H] 
\centering 
\subfloat[$p=50$]{\includegraphics[width = 0.8in]{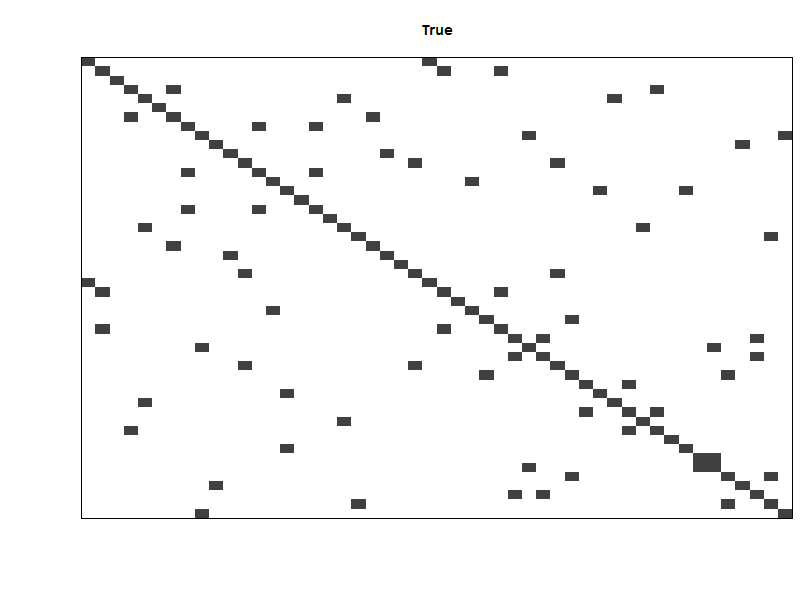} 
\includegraphics[width = 0.8in]{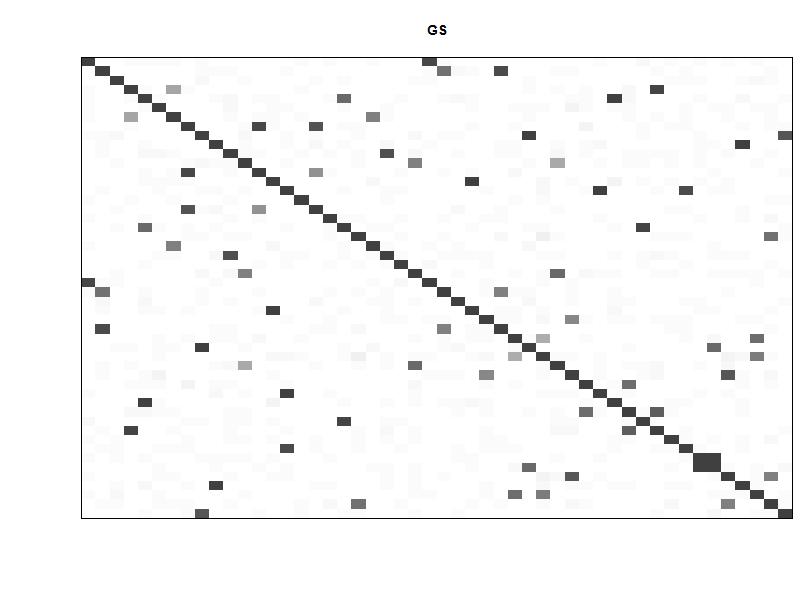} 
\includegraphics[width = 0.8in]{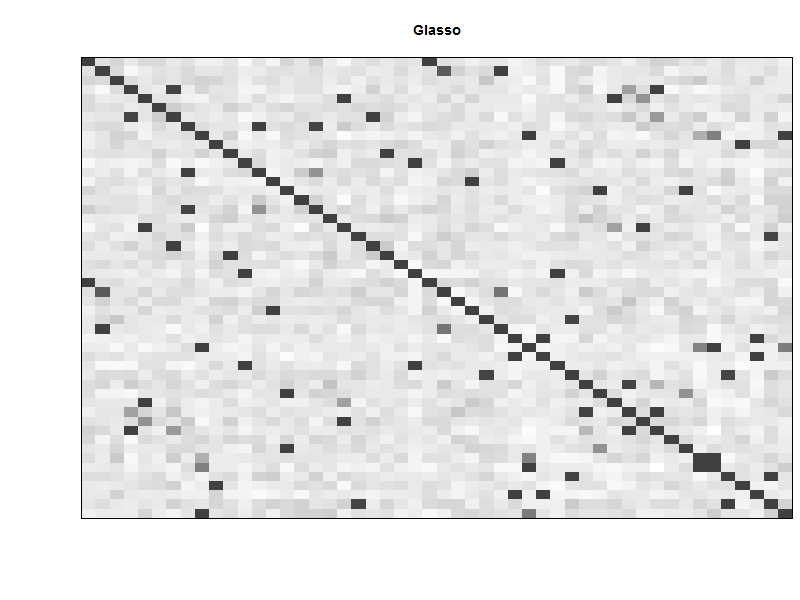}} 
\includegraphics[width = 0.8in]{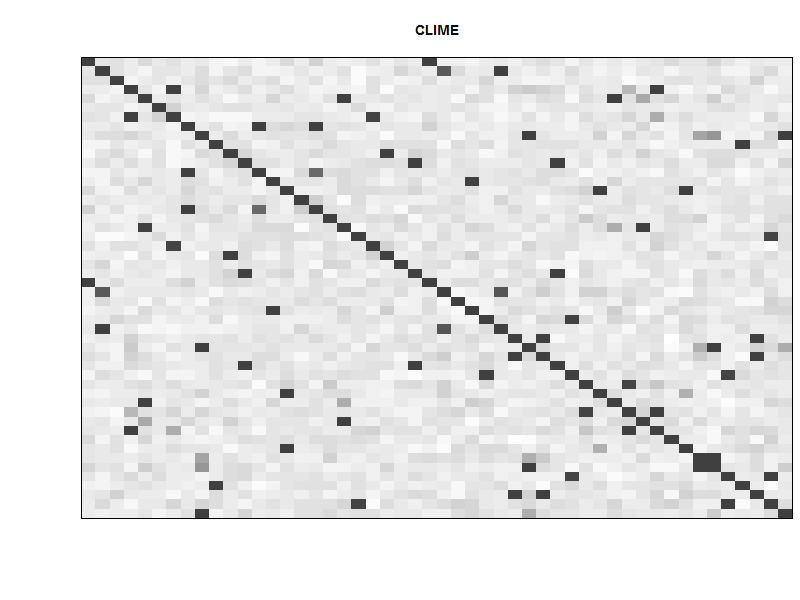} \\ 
\subfloat[$p=100$]{\includegraphics[width = 0.8in]{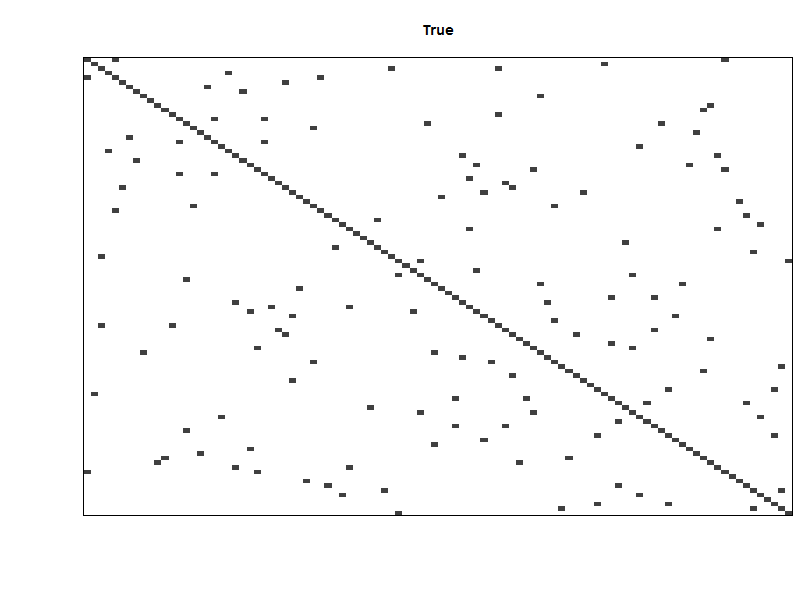}
\includegraphics[width = 0.8in]{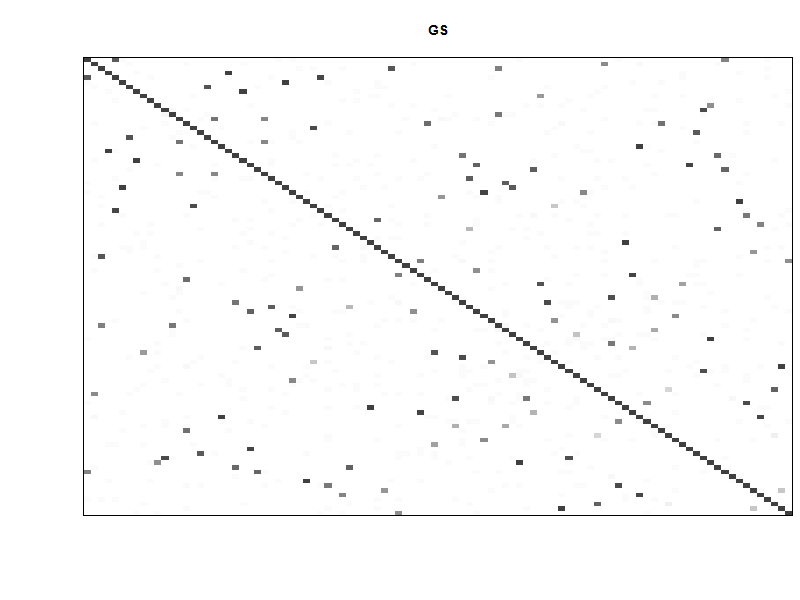}
\includegraphics[width = 0.8in]{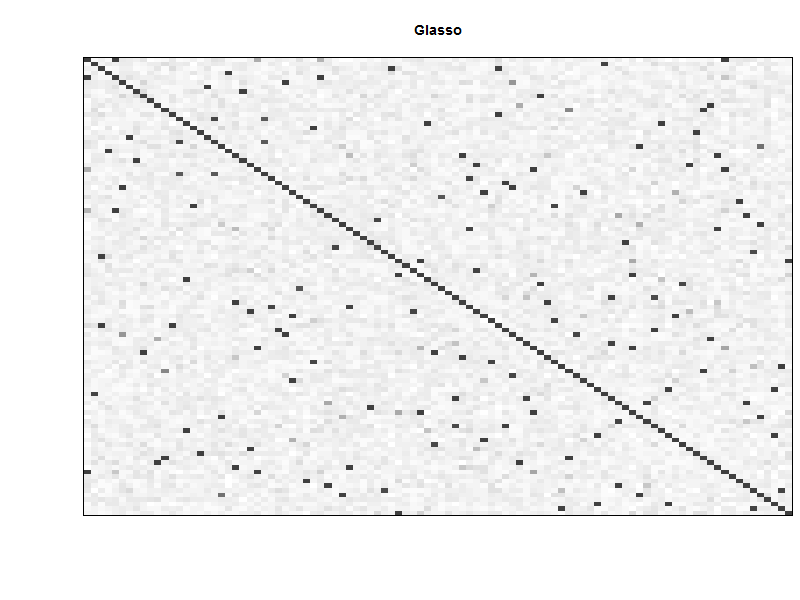}}
\includegraphics[width = 0.8in]{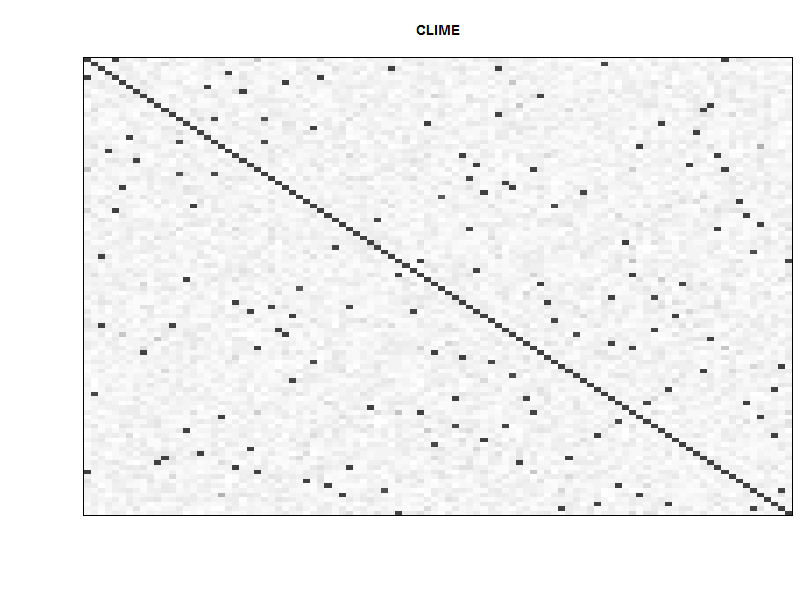} \\
\subfloat[$p=150$]{\includegraphics[width = 0.8in]{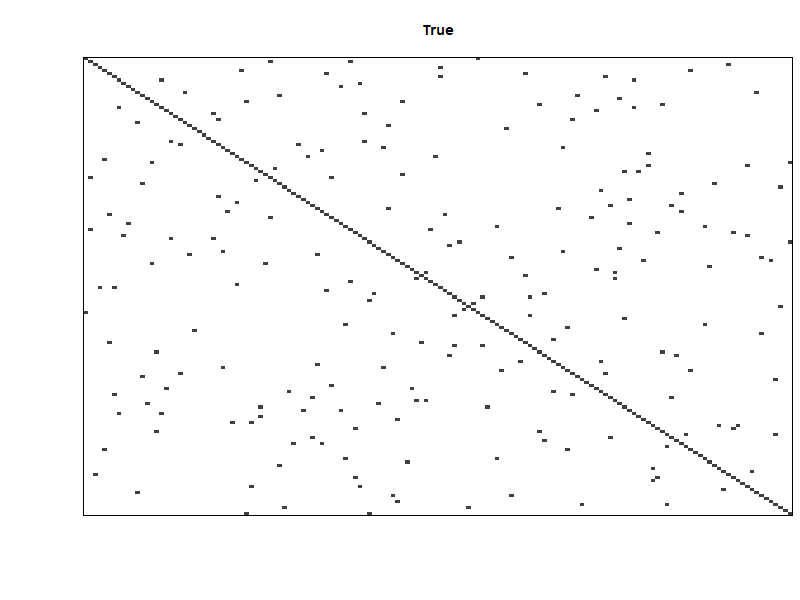}
\includegraphics[width = 0.8in]{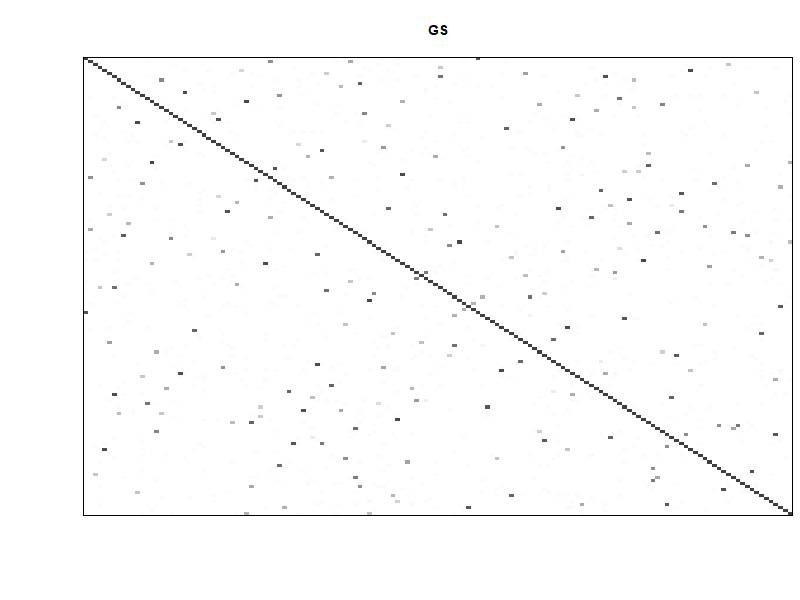}
\includegraphics[width = 0.8in]{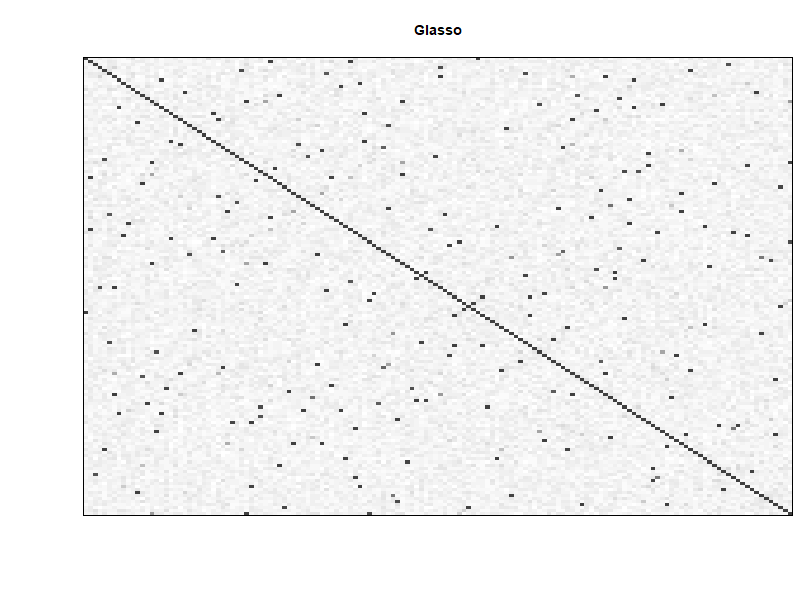}}
\includegraphics[width = 0.8in]{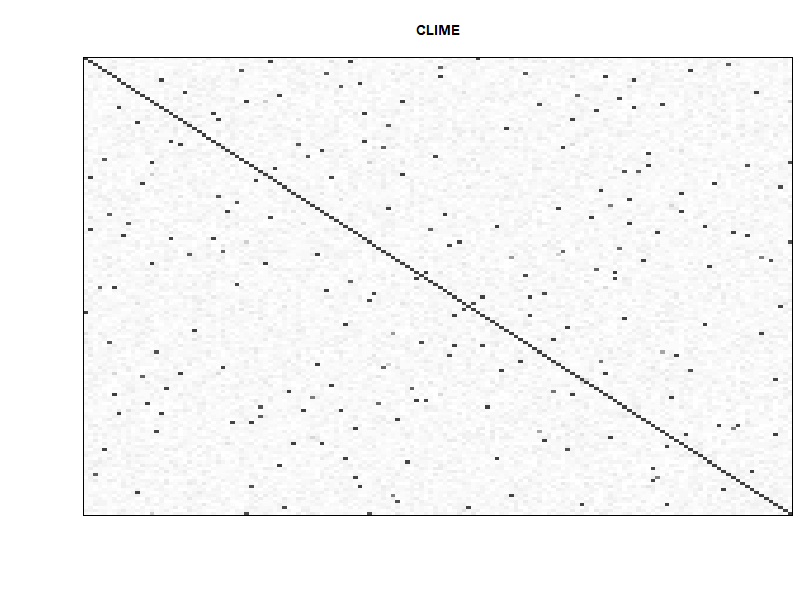} \\
\caption{Model  $\text{NN}(2)$.  Heatmaps of the frequency of the zeros identified for each entry of the
precision matrix out of $R=50$ replications. White color is 50 zeros
identified out of 50 runs, and black is 0/50.}
\label{Heatmaps_2}
\end{figure}

\begin{figure}[H]
\centering
\subfloat[$p=50$]{\includegraphics[width = 0.8in]{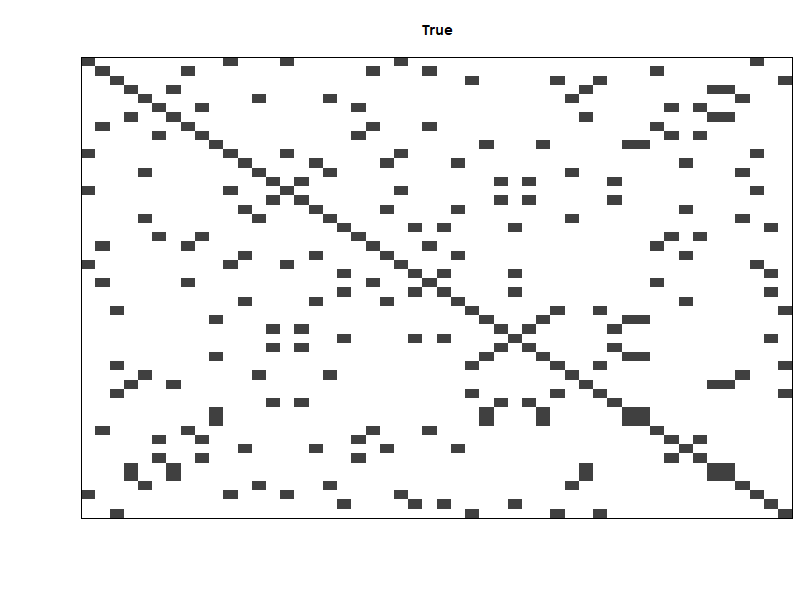}
\includegraphics[width = 0.8in]{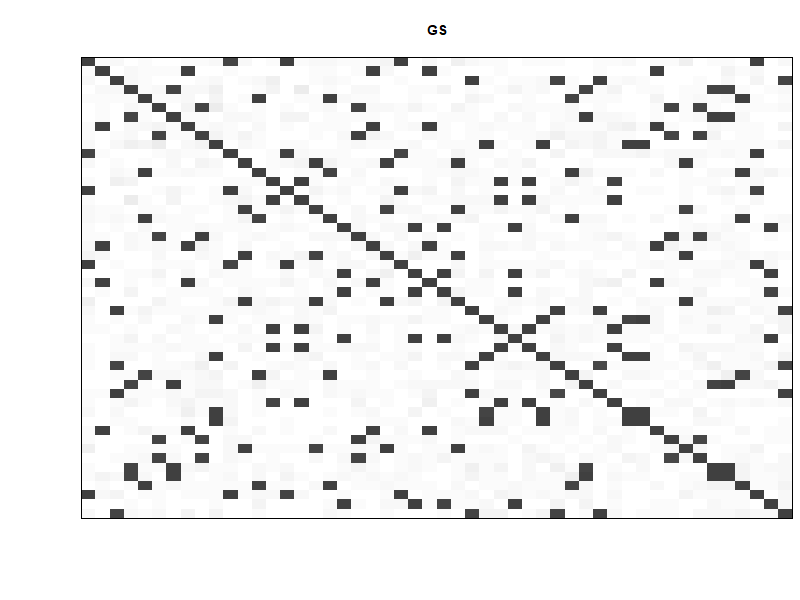}
\includegraphics[width = 0.8in]{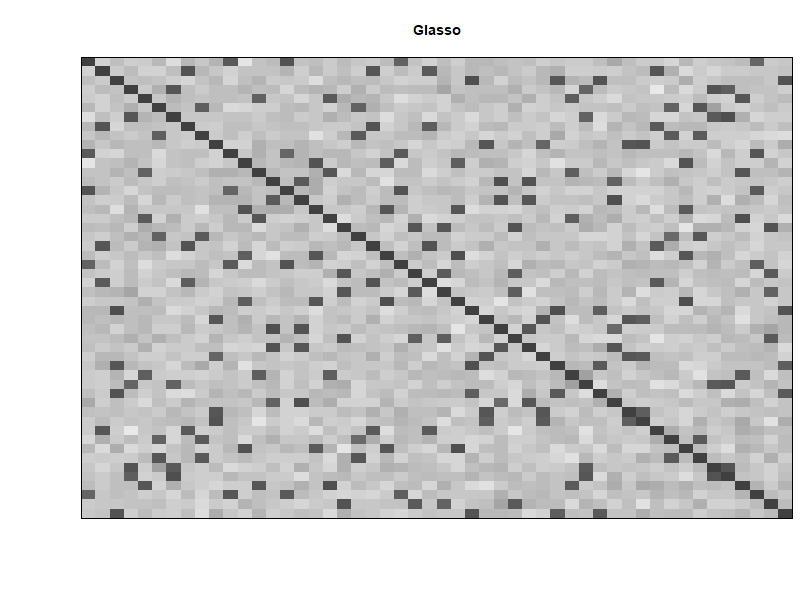}}
\includegraphics[width = 0.8in]{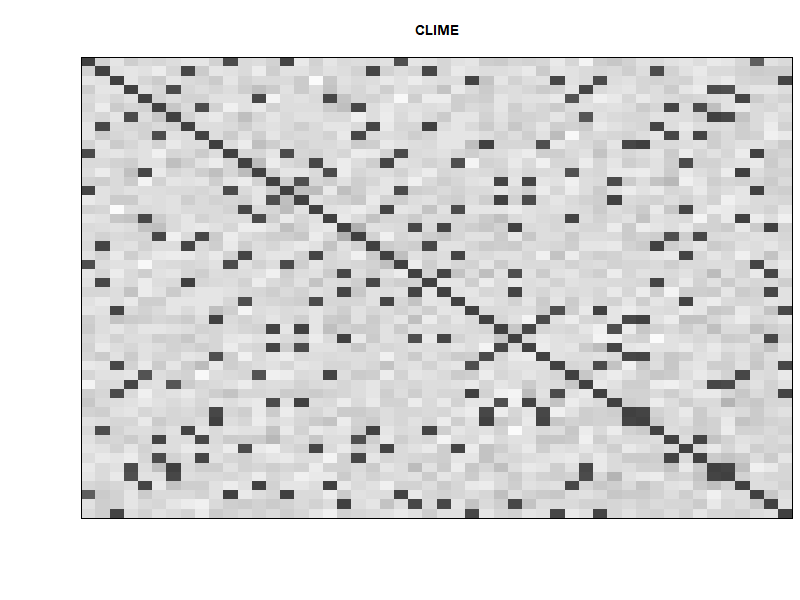} \\
\subfloat[$p=100$]{\includegraphics[width = 0.8in]{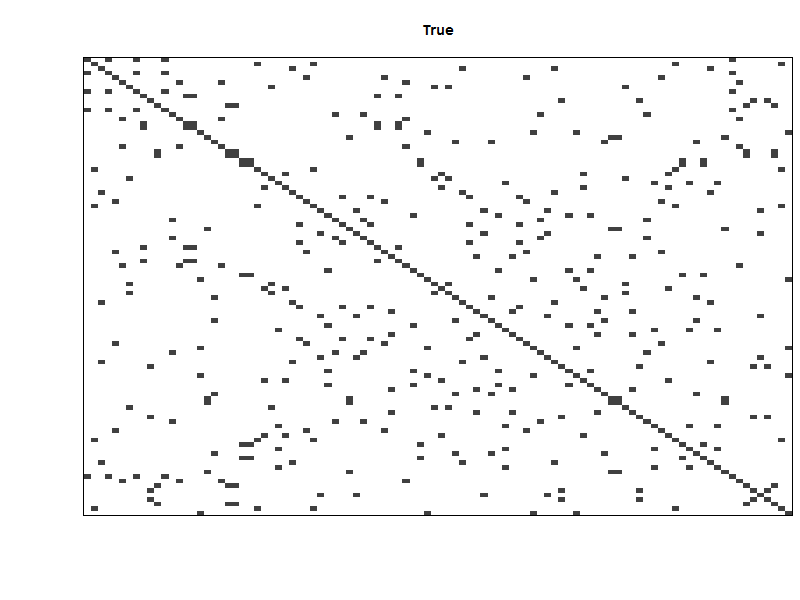}
\includegraphics[width = 0.8in]{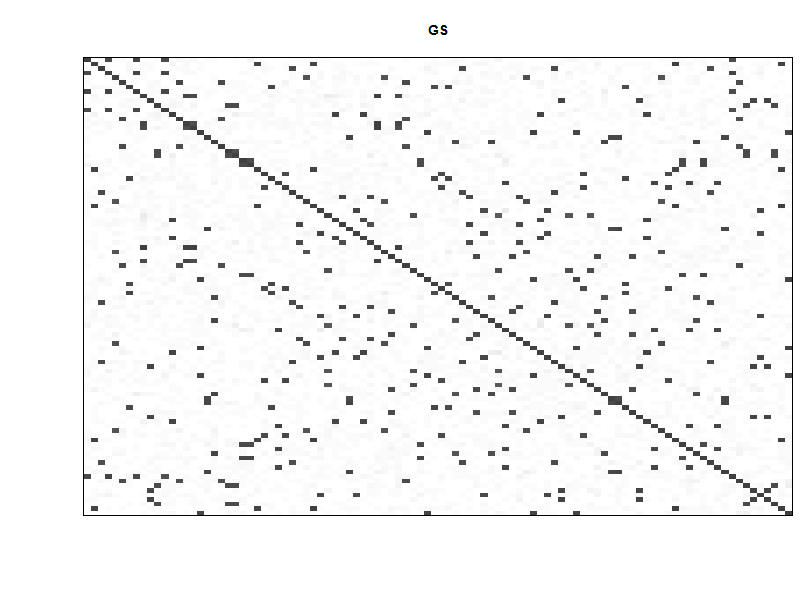}
\includegraphics[width = 0.8in]{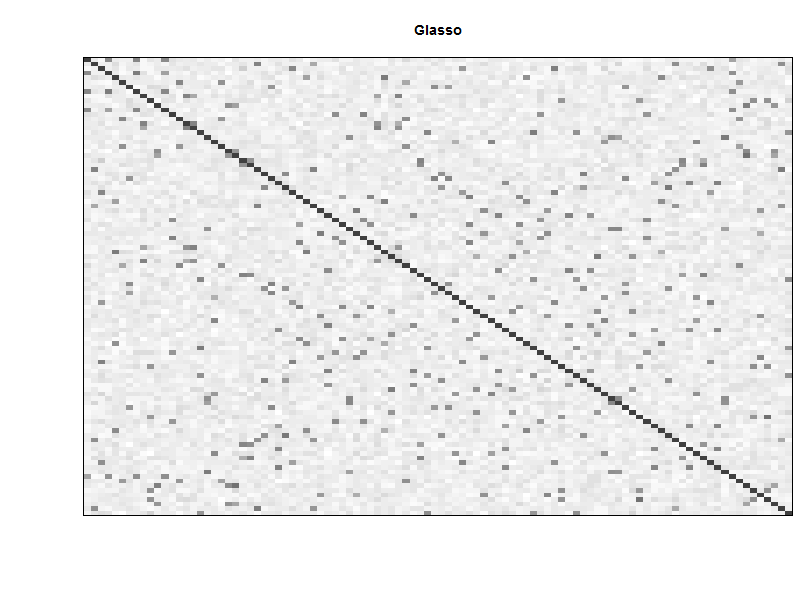}}
\includegraphics[width = 0.8in]{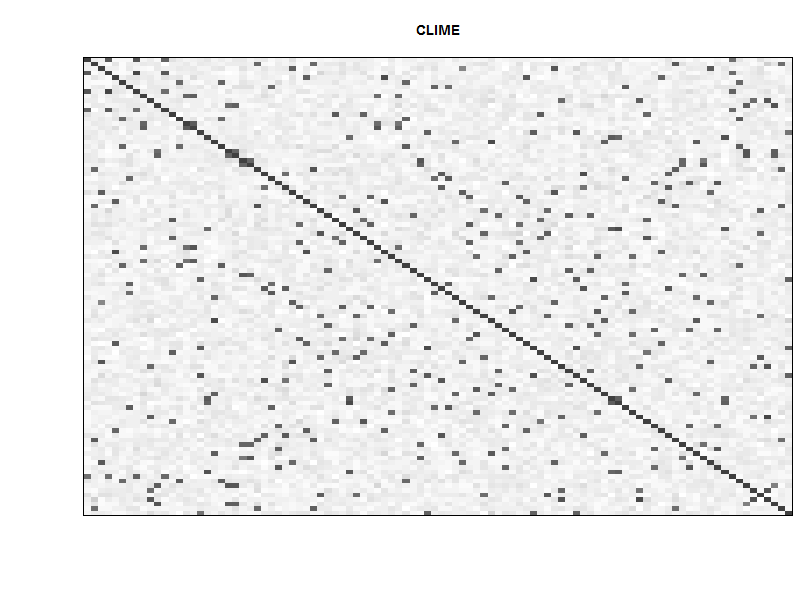} \\
\subfloat[$p=150$]{\includegraphics[width = 0.8in]{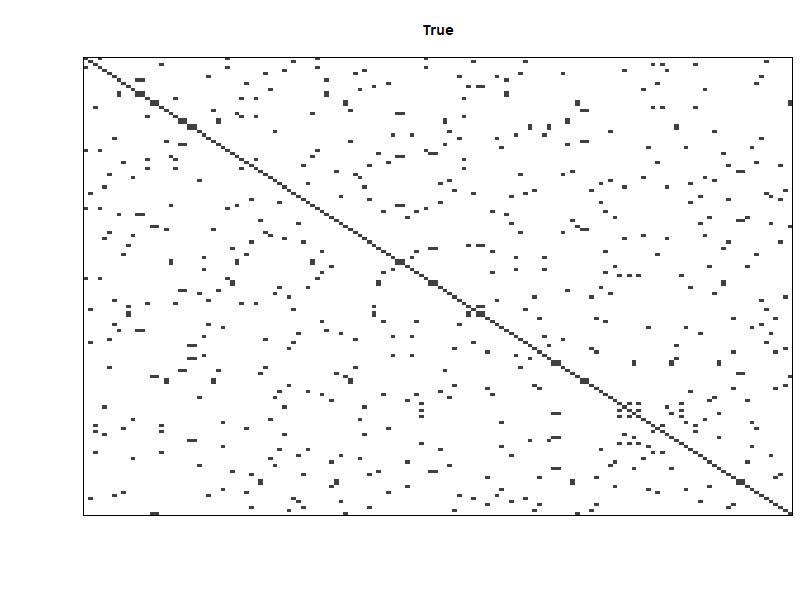}
\includegraphics[width = 0.8in]{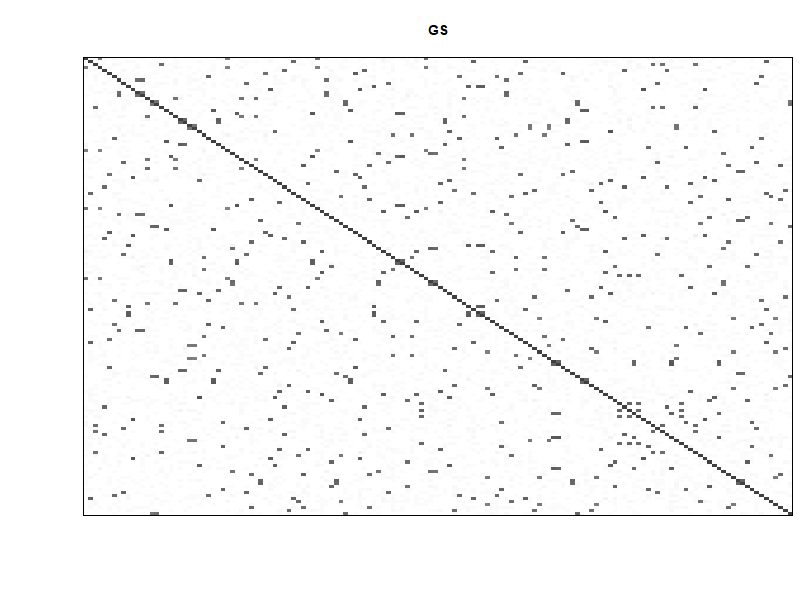}
\includegraphics[width = 0.8in]{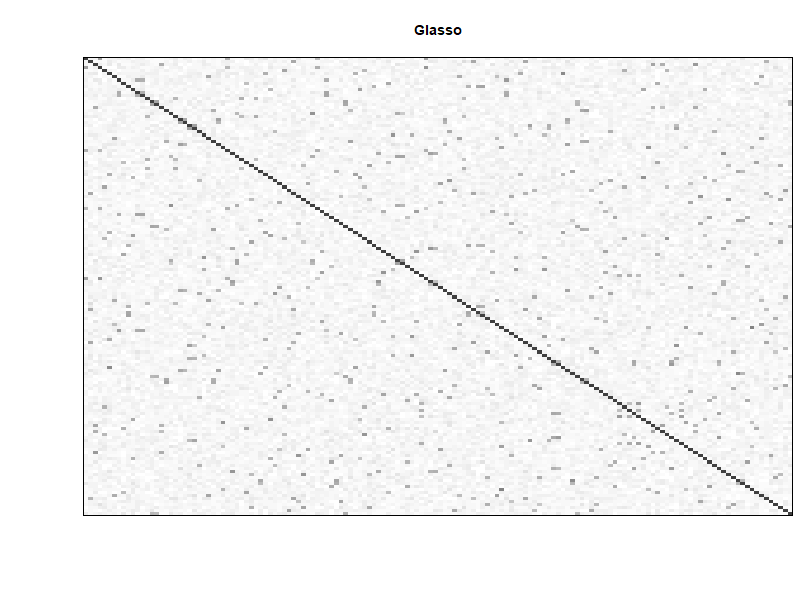}}
\includegraphics[width = 0.8in]{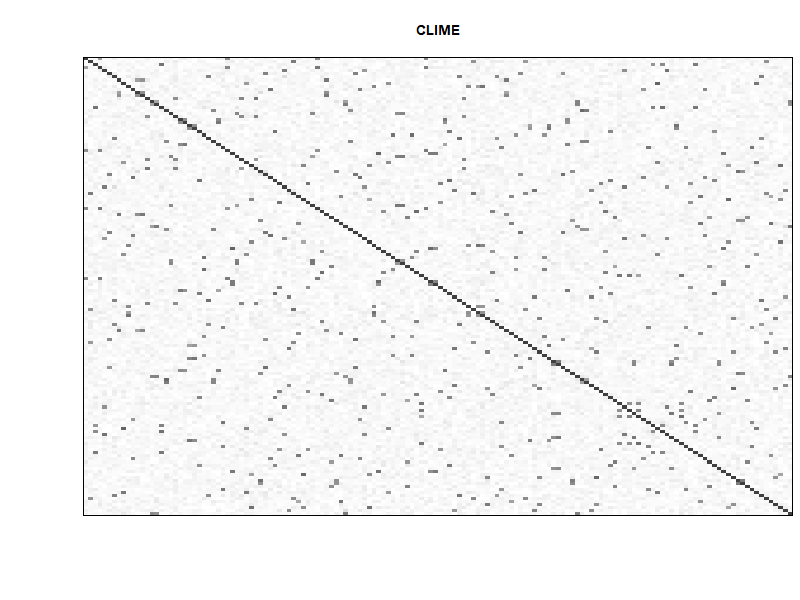} \\
\caption{Model  BG.  Heatmaps of the frequency of the zeros identified for each entry of the
precision matrix out of $R=50$ replications. White color is 50 zeros
videntified out of 50 runs, and black is 0/50.}
\label{Heatmaps_3}
\end{figure}

\bibliography{bibliography}

\bibliographystyle{chicago}

\end{document}